# Developing Competency in AI Ethics: A training framework

**Alexandra Prégent[1] & Alexei Grinbaum[1]**

[1]CEA-Saclay, 91191 Gif-sur-Yvette, France

**Abstract**

We present a training framework for developing core competencies in the practical operationalisation of AI ethics: (1) identifying ethical concerns; (2) determining which standard ethics requirements they implicate; (3) selecting and prioritising concrete technical and organisational measures to address them; and (4) recognising ethics tensions and reasoning about trade-offs where such tensions cannot be fully resolved. The framework is intended for researchers and engineers in AI, Research Ethics Committee members, policymakers, and students, providing them with practical tools for translating abstract ethical requirements into concrete actions. The framework provides a step-by-step approach organised around four interconnected tasks: issue identification, requirement mapping, measure selection, and tension identification. These tasks are supported by a suite of training materials, including an Ethics Checklist and a Portfolio of Measures. Here, we complement them with a set of practical exercises of three different types as well as trainer guides. We also discuss lessons learned in pilot training sessions and offer recommendations for trainers.

## 1. Introduction

Knowing that fairness matters does not tell an engineer *how* to design a fair AI system. Knowing that human oversight matters does not determine what form of oversight should be implemented, by whom, or at which stages of an AI system's operation. Similarly, recognising the importance of transparency does not specify what information should be disclosed, to whom, or under what conditions. With hundreds of ethics guidelines now available (Corrêa et al. 2023; Jobin, Ienca, & Vayena 2019; Hagendorff 2020; Fjeld et al. 2020), AI ethics researchers are trying to bridge the gap between high-level ethics principles and concrete implementation measures (Hallensleben et al. 2020; Morley et al. 2020; Saltz & Dewar 2019). Practitioners from diverse fields including researchers and engineers in AI, Research Ethics Committee (REC) members, policymakers, and students are expected not only to identify ethical concerns, but also to determine how these concerns should be addressed in concrete socio-technical contexts.

A recent survey (Evans & Maris 2026) revealed uneven confidence across professional tasks, with particular difficulties concerning the translation of ethics requirements into design measures. In Figure 1, we synthesise a set of three common desiderata with regard to training in AI ethics, as expressed by survey participants.

| **Desideratum 1**: Application-oriented training (Learning-by-Doing). | **Desideratum 2**: Realistic and sufficiently complex use cases. | **Desideratum 3**: Concrete tools for applying ethics principles in practice. |
|---|---|---|

*Figure 1. Desiderata for training in AI ethics based on Evans and Maris (2026).*

In response to these desiderata, we propose a training framework to develop competency in operational AI ethics in four steps: 1) identify which features of an AI system or project raise ethical concerns; 2) determine which broader ethics requirements are implicated by these concerns and why; 3) consider and justify concrete technical or organisational measures that could address them; finally, because ethics requirements and the measures proposed to address them may conflict, practitioners must be able to 4) recognise and reason through tensions between competing ethics objectives, identifying cross-cutting measures where possible and justifying the resulting trade-offs where conflicts cannot be fully reconciled. The framework is thus organised around these four interconnected competencies, by transforming them into tasks that we will respectively refer to as: issue identification, requirement mapping, measure selection, and tension identification. These tasks are accompanied by practical training materials including an Ethics Checklist, a Portfolio of Measures, three different types of case-based exercises (ethical dilemmas, mock review exercises, and ethics-by-design exercises), and Companion documents, tailored to each exercise (Figure 2).

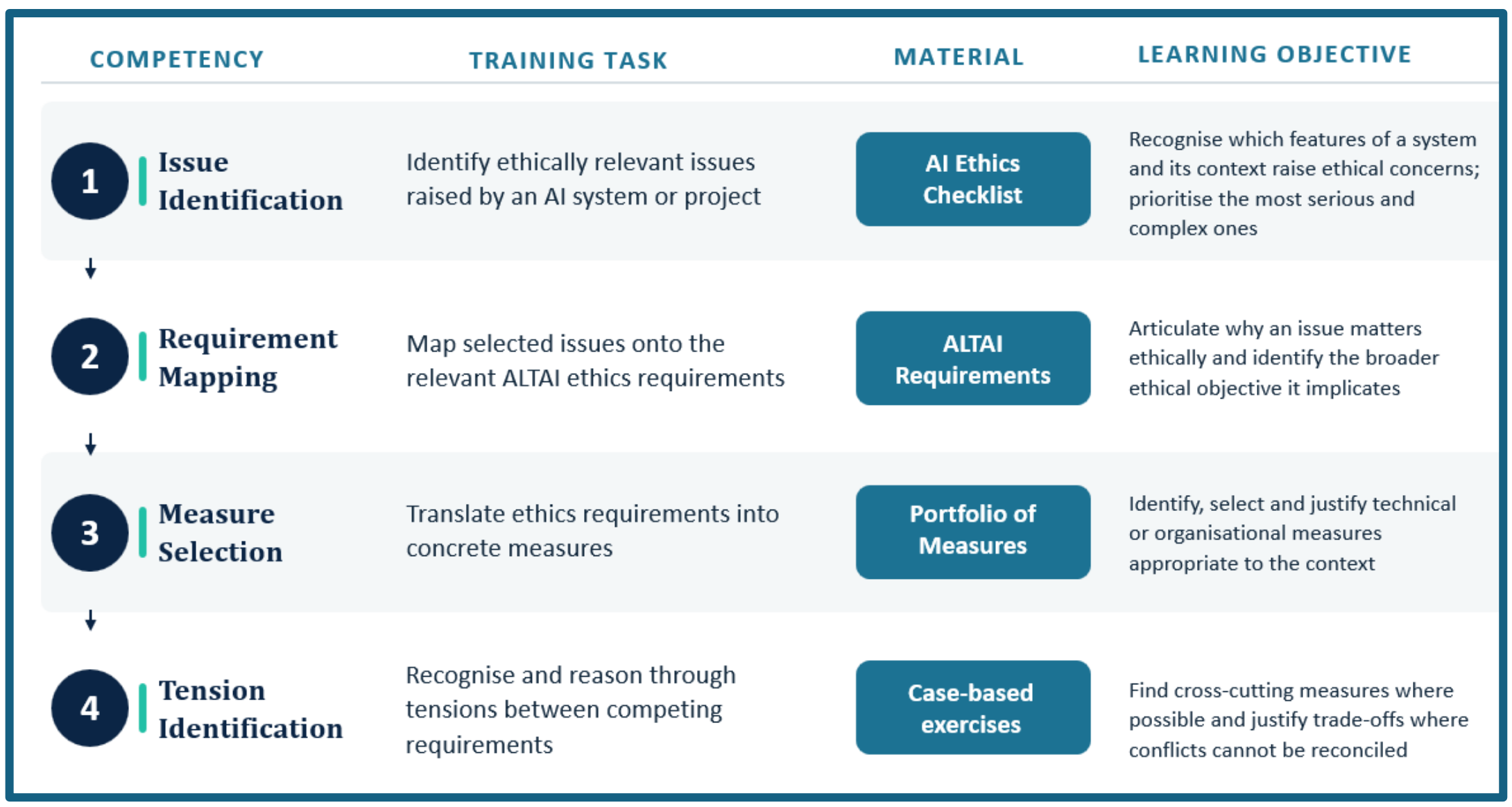

*Figure 2. Targeted competencies and associated training task, material, and learning objectives*

The content of the training materials was built using two resources. The first is the Assessment List for Trustworthy AI (hereinafter ALTAI), developed in 2019-2020 by the High-Level Expert Group on AI (European Commission 2020). ALTAI provides the higher-level ethics requirements that we use to structure the identification and mapping of ethical issues. The second resource is bottom-up co-creation activity in the Horizon Europe project AIOLIA, in which academic and industrial partners translated AI ethics requirements into implementation measures (Bayerl et al. 2026; Aires et al. 2026). This was conducted on six different European use cases, providing the basis for constructing realistic training scenarios. This approach responds to the need for authentic, context-rich cases for experiential learning and applied ethical reasoning (Bombaerts et al. 2021; Corrêa et al. 2025; Knight et al. 2025). It provides an explicit learning sequence from ethical concerns to implementation choices; in this, it differs from our previous work (Aucouturier & Grinbaum 2025a) as it adds the Portfolio of Measures, an explicit mapping of ethical issues under ALTAI requirements, a systematic analysis of tensions, and case-based exercises beyond healthcare.

Moreover, to accommodate different audiences, we developed two broad versions of the training materials. One version ("Reviewer path") centres on the task of *reviewing* an AI system or project and primarily targets European ethics appraisal scheme experts, AI governance and policy experts, non-AI researchers, and students. The second version ("Designer path") centres on the task of *design and development* choices, targeting AI researchers, engineers, and developers, including both experts and students.

The remainder of this paper is structured as follows. **Section 2** explains the framework step-by-step and introduces the training materials. **Section 3** illustrates how this framework can be used in practice through our set of practical exercises. **Section 4** then discusses lessons learned and challenges encountered during pilot trainings and provides guidance for trainers. **Section 5** concludes.

## 2. The framework step-by-step

The following subsections describe the four stages of the framework and introduce the related training materials.

### 2.1 Identifying ethics issues: the AI Ethics Checklist

The first competency concerns the ability to identify ethically relevant issues arising from an AI system or project. Before practitioners can determine which ethics requirements are implicated or consider how those requirements might be operationalised, they must be able to recognise which features of the system and/or its wider socio-technical context raise ethical concerns. For instance, the use of a surveillance camera in a prison, at an airport, or in a school does not lead to the same practical ethics measures.

To support the development of this competency, participants are provided with an AI Ethics Issue Identification Checklist (see Annex 1). This Checklist is derived from Aucouturier and Grinbaum's (2023, 2025a) Checklist and is structured around the seven ALTAI requirements: human agency and oversight, technical robustness and safety, privacy and data governance, transparency and explainability, non-discrimination and fairness, societal and environmental well-being, and accountability. The Checklist associates these high-level requirements with questions that would potentially help to identify relevant issues. We use the Checklist for a limited purpose: to help practitioners overcome the "where to begin" problem that often arises when they face an ethically complex project. The Checklist is thus designed as a tool to support the development of the first competency, namely the ability to identify ethically relevant issues in an AI system or project. Accordingly, it is deliberately non-exhaustive and not intended to determine, by itself, which issues are ethically relevant in a given case. The Checklist is intended to remind practitioners of the range of issues that should not be forgotten, not to provide a fixed inventory of risks that participants must identify. Practitioners are expected to interpret, adapt, and supplement the proposed questions in light of the specific system and its socio-technical context.

The use of checklists has been questioned in the literature (Kinoshita et al. 2025; Kijewski, Ronchi, & Vayena 2025; Bolte, Vandemeulebroucke, & van Wynsberghe 2022; Mittelstadt 2019), and we have recently provided some answers to these criticisms (Aucouturier & Grinbaum 2025b). Critics have, among other things, pointed out that ethical judgement cannot be reduced to a fixed set of questions or procedures, since the relevance and meaning of ethics requirements depend on context (Mittelstadt 2019). There is also a risk that such tools encourage a "tick-box" approach to ethics, where practitioners are under the impression that "completing" the checklist's questions guarantees de facto compliance with the given ethics requirement (Madaio et al. 2020; Morley et al. 2020; Morley et al. 2021).

In response, we explicitly encourage discussion and debate during the use of the Checklist in group exercises, providing space and time for ethical reflection. Because real-world scenarios are complex and bring entangled context-specific concerns, participants are not expected to identify every conceivable ethical issue during the training. They are, on the contrary, encouraged to distinguish serious and complex issues from less urgent ones through dialogue and reflection. They are expected to exercise critical judgement as motivated, among other reasons, by considerations of feasibility: in real life, operational AI ethics necessarily takes place under practical constraints, meaning that researchers and engineers will not be able to address every potential concern with equal depth or speed. The task of issue identification therefore also constitutes an exercise in determining which aspects of an AI system warrant more urgent analysis.

### 2.2 Mapping ethics issues under ALTAI requirements

The second competency concerns the ability to relate concrete ethical issues to the standard ethics requirements that they implicate. As discussed in Section 4, while this step may seem intuitive at first sight, early pilots revealed that participants had difficulties correlating identified ethics issues with a relevant ethics requirement. Requirement mapping helps participants articulate, sometimes in many alternative ways, *why* a particular issue matters ethically and what broader requirement this particular issue implicates. Hence, once participants have identified and prioritised ethically relevant issues, they are asked to map them onto the relevant ALTAI requirements. This also helps participants move beyond isolated problems and examine how ethics issues interact at both the level of system design and the contextual one.

For example, concerns about automated reporting in healthcare may relate to the standard ALTAI requirement of human agency and oversight, or to the requirement of technical robustness and safety. Similarly, concerns raised by behavioural profiling in the workplace may be classified as related to the requirements of privacy, fairness, transparency, or human agency. The relevant mapping will depend on the characteristics and purpose of the system, the context in which it is deployed, as well as on the subjective judgement of the evaluator. The use of ALTAI at this stage is meant to provide a common normative vocabulary across different cases and professional contexts.

### 2.3 Selecting measures: The Portfolio

The third competency constitutes the operational core of the framework and concerns the ability to translate ethics requirements into concrete measures. Once participants have identified ethically relevant issues and mapped them onto the relevant ethics requirements, they must be able to determine what could actually be done or changed to address those concerns. This step responds directly to a recurrent difficulty reported by practitioners who may feel rather confident in their capacity to recognise relevant ethics requirements or concerns, but they often remain uncertain about *how* to address them in a given scenario (see Desideratum 3 in Figure 1).

To support this task, we use a Portfolio of Measures (see Annex 2), which is organised around the ALTAI requirements for consistency. The Portfolio proposes measures that are categorised as "organisational," "technical," or "both". Organisational measures primarily focus on how the *provider* of an AI model or system incorporates and manages responsible AI design. These measures thus target organisational practices, structures, policies, standard operating procedures, and governance frameworks. Technical measures primarily focus on the *design* aspect of AI models or systems. They target specific techniques, routines, processes, or design features. The category "both" refers to measures that cover both technical and organisational components. For instance, Measure 27 of our portfolio is categorised as "both" because it proposes that the organisation addresses "unacceptable manipulative practices through corrective actions, design changes, or restrictions on AI use", which requires both technical and organisational efforts.

The Portfolio of Measures was built through AIOLIA's bottom-up approach to operationalising AI ethics and is the result of three different initiatives: (1) a co-creation process involving academic, policy, and industry partners who proposed measures for real-world use cases (Bayerl et al. 2026); (2) several rounds of internal review and adaptation; and (3) pilot testing during training sessions. While the Portfolio broadly follows the ALTAI requirements, we also divided some of them into sub-requirements to achieve more targeted measures and facilitate the translation task. For example, ALTAI requirement #1, named "Human Agency and Oversight", was divided into sub-requirement #1a, named "Oversight", and sub-requirement #1b, named "Human Autonomy and Agency" (see Annex 2).

Similarly to the Checklist, the Portfolio of Measures is intended to provide participants with a repertoire of possible practical measures, but it is non-exhaustive and not presented as a catalogue of "correct solutions". Its purpose is to support practitioners' reasoning by offering suggestions that are clearly linked to ethics requirements. Participants must consider, preferably in dialogue with each other, which measures are appropriate to the particular system and context, what ethics requirement each measure is intended to support, and what limitations or unintended consequences it may involve. Importantly, the relationship between a measure and an ethics requirement is *not* fixed and overlap is explicitly acknowledged. The appropriateness and effectiveness of a measure depend on the characteristics of the AI system, its intended use, the people affected by it, the organisational setting, and the resources and constraints under which it operates. A measure that is appropriate in one socio-technical context may be ineffective, disproportionate, or impracticable in another. Participants are therefore encouraged to treat the Portfolio as a repertoire of possibilities that must be evaluated, adapted, and expanded as needed. They are also encouraged to formulate their own targeted measures.

Measures selected from the Portfolio must be prioritised. Participants may identify several possible measures addressing the same requirement, but these measures may differ in their effectiveness, feasibility, costs, implementation burden, or effects on other requirements. Participants are invited to discuss and justify their selection, further developing their ability to reason about not only whether a measure is ethically relevant, but also why it should be preferred over alternative responses in the particular context at hand.

### 2.4 Identifying potential ethics tensions

The fourth competency concerns the ability to recognise and reason through tensions between ethics requirements. AI ethics rarely involves only compatible objectives: when in tension, operationalising one requirement can compromise the operationalisation of another, because the measures introduced to address one undercut the measures required to instantiate another. For example, care or protection measures that instantiate the requirement of human safety often clash with measures introduced to respect human autonomy and agency. Identifying and addressing these tensions is a key exercise to enhance the participants' ethical reflection.

The purpose is not simply to identify situations in which two requirements appear to conflict, but to examine how changing the set of measures may strengthen one requirement while weakening another and how competing objectives may constrain the range of feasibility with regard to the selected measures. For example, strengthening transparency through audit logs improves accountability while creating privacy risks; monitoring the chain-of-thought output of an AI model may potentially create cybersecurity concerns.

Where a measure appears to significantly undermine another requirement, participants are asked to consider alternative measures or combinations of measures. Particular attention is given to the identification of cross-cutting measures that may support several requirements simultaneously, thereby reducing the cost of the trade-offs involved. At the same time, participants are not expected to eliminate all tensions. The practical exercises that we designed are based on real and complex use cases where, as in real life, some competing objectives will remain in tension. This task therefore places particular emphasis on justification and prioritisation. Participants are asked to explain which requirements they consider most important in a particular context, why they regard trade-offs as acceptable or unacceptable, and how their proposed measures respond to the tensions.

## 3. Learning-by-Doing: Practical exercises

The use of our framework is based on practical exercises. This section introduces the exercises and explains how they are designed to enable participants to apply the framework.

We use the format of three-hour online and in-person training sessions. The composition of the sessions is different for the Reviewer and the Designer paths (Figure 3). In both versions, the structure of the training includes a theory session (30 minutes), a practical exercise (90 minutes) followed by a short break, and a plenary discussion (45-60 minutes). In a joint theory session, facilitators deliver only a short and straightforward explanation of the work step-by-step. They do not teach AI ethics from scratch but rely on the existing knowledge of the participants, who are asked to prepare for the session in advance. Rather, the goal is to introduce the participants to the Ethics Checklist and the Portfolio of Measures. The facilitators also emphasise that ethics analysis is deeply contextual and illustrate the concept of ethics tensions and practical trade-offs on standard examples.

In the practical exercise part, we divide participants into small groups of up to seven people. Participants first familiarise themselves with the exercise. They are then asked to identify ethics issues and to select the most serious and complex ones, to map the selected issues to the AI ethics requirements, and to select potential practical measures. If time remains, participants are invited to discuss tensions between ethics requirements and readjust the selected measures accordingly.

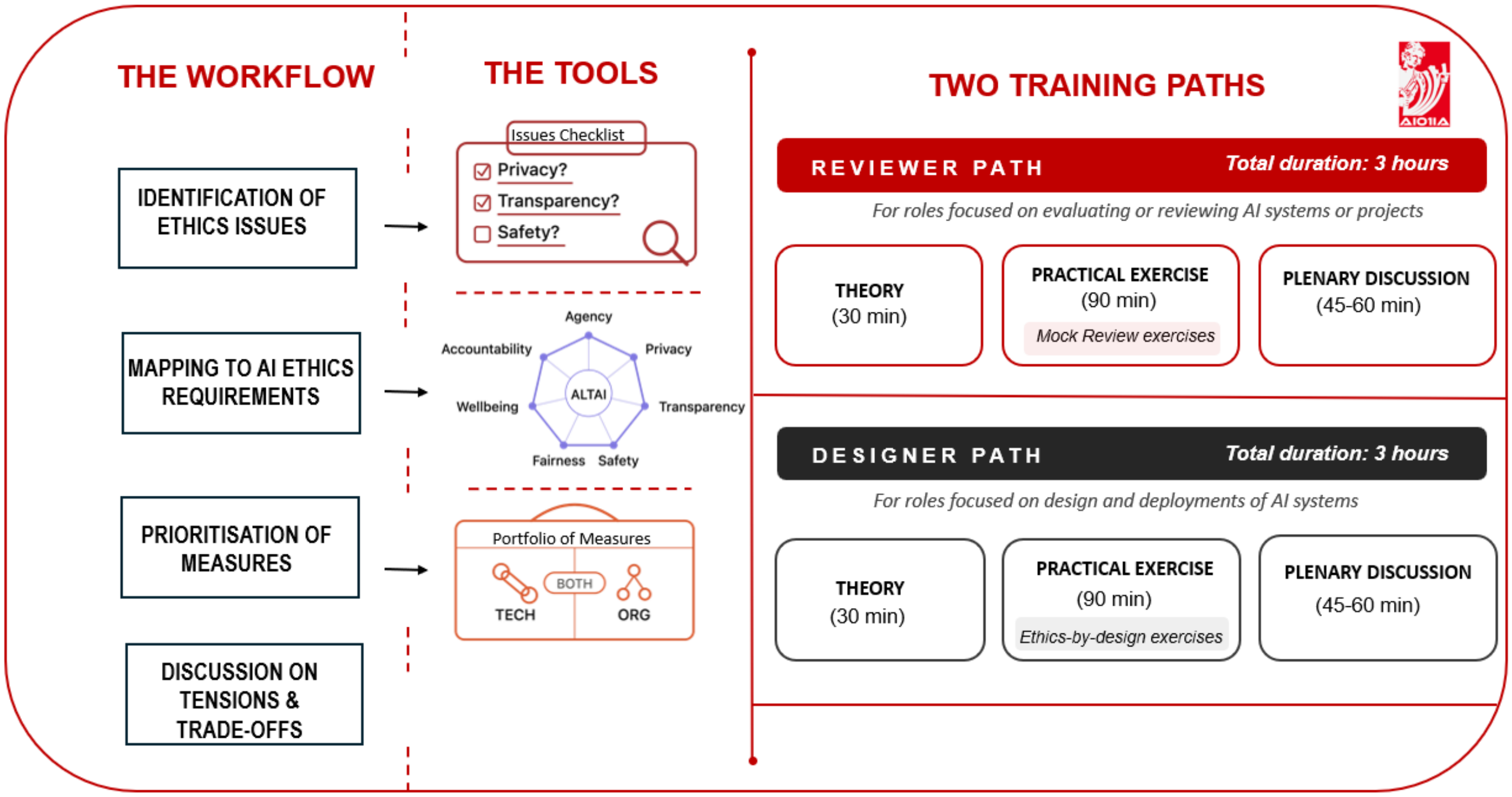

*Figure 3. Two training paths for Reviewers and Designers.*

The final plenary discussion is intended for a presentation of the exercises by each group. Participants explain their interpretations and choices to the other groups and compare competing interpretations if several groups were working on the same exercise.

Because the practical exercise is quite demanding, we also propose a "Beginner" version, where the theory part is extended to one hour and covers foundations of AI ethics, and where participants are not asked to select practical measures but only to identify issues and map them to relevant ethics requirements.

Our practical exercises aim at helping a wide range of participants; hence, we developed three different exercise types on which they can apply this framework: ethical dilemmas, mock reviews, and ethics-by-design practical exercises. Mock review exercises have been developed for the Reviewer path and are meant to guide participants in the evaluation of a given AI project. Ethics-by-design exercises have been developed for the Designer path, and are meant to guide participants in the concrete development and/or deployment of an AI system. We also created ethical dilemmas, which can be used either in the practical exercise part for Beginner audiences, or in the theory part, as a way to explain the broader idea behind the more demanding exercises.

Each exercise presents participants with realistic scenarios taken from AIOLIA's use cases, and requires them to apply the framework in practice. Below we present one example of each type. More examples are given in the Annexes.

### 3.1 Ethical dilemma exercises

The ethical dilemmas (Annex 3) are easy-to-use exercises mostly intended for the "Beginner" version of the training. Their main pedagogical objective is to develop a sense of seriousness and difficulty with regard to AI ethics requirements, based on short scenarios. Depending on the exercise chosen, learners develop the ability to (1) recognise the presence of ethics requirements at play in a given scenario, (2) realise that competing AI ethics requirements may clash against each other, and (3) understand that justification is key to ethical reasoning.

Dilemma exercises introduce participants to the contextual interpretation of AI ethics requirements through guided discussion. Each exercise presents a situation in which multiple ethics requirements appear legitimate but cannot be satisfied simultaneously. Rather than searching for a single "correct solution," participants are encouraged to identify the competing principles and values, discuss conflicting views and positions, and justify their reasoning along the way.

For example, the *Human-AI Interaction* ethical dilemma (Figure 4; see also Annex 3) takes the form of a fictional but realistic-looking dialogue to illustrate the tension between the requirements of privacy and safety. The core of the tension is in the request from the human for privacy during a conversation that involves a safety risk, because the user reveals that they are "panicking". Participants are invited to identify the tension and to reflect on whether the bot should send a safety alert to a third party (parent, teacher, doctor) despite its commitment to protecting privacy. An additional level of tension is added with regard to transparency, when the facilitator asks whether the chatbot, in the event it decides to send an alert to a third party, should also disclose this information to the user.

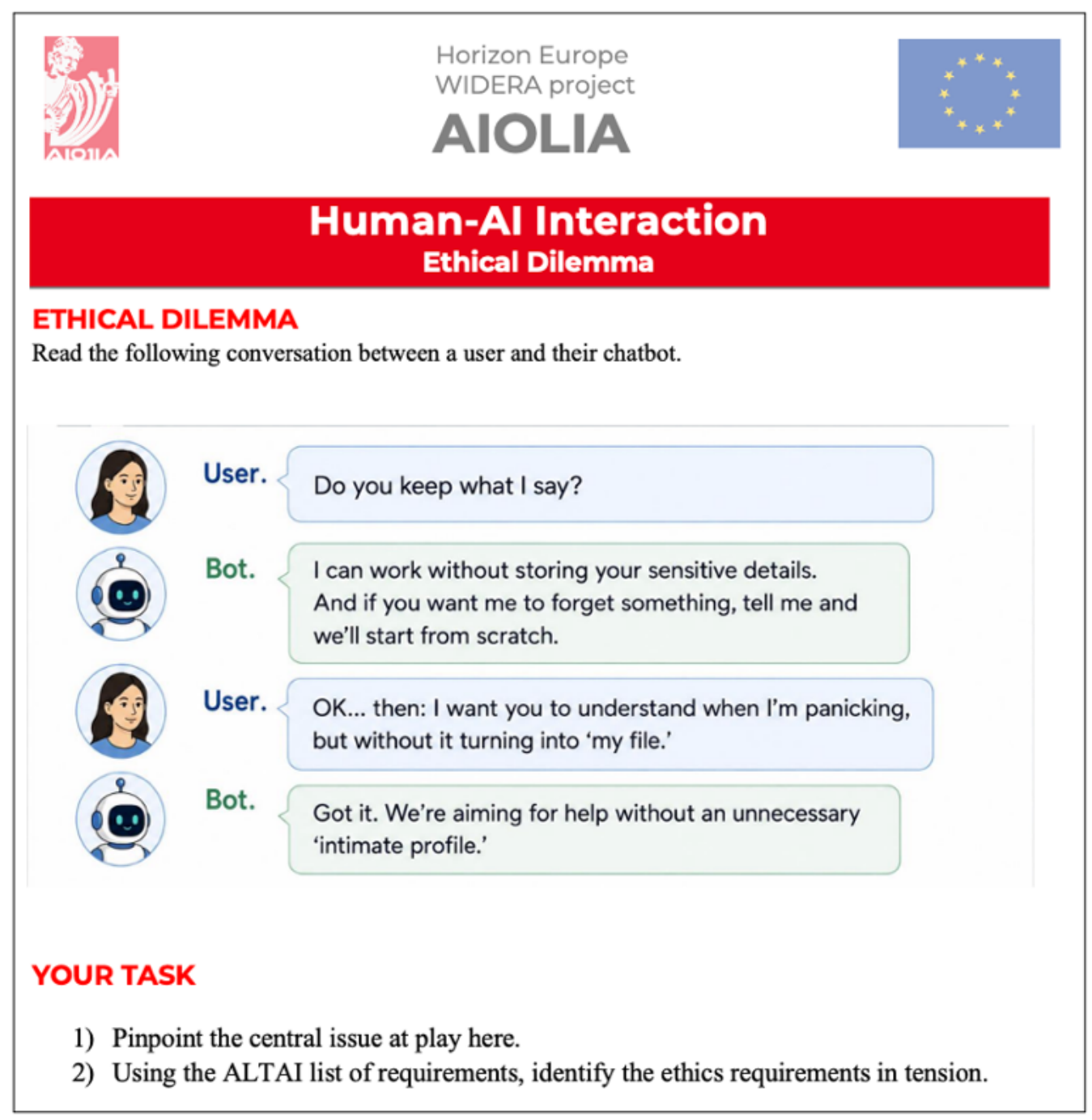

*Figure 4. Ethical dilemma "Human-AI Interaction".*

### 3.2 Mock review exercises

The mock review exercises (Annex 4) are intended to develop professional operational competencies related to ethics assessment and AI governance. They develop the ability to propose concrete solutions in AI design rather than simply identifying ethical problems. These exercises place participants in the role of ethics reviewers for fictional but realistic-looking AI research and development projects. They first present a project description containing limited but controversial technical and organisational information as well as an example of use. Participants are then asked: (1) to use the Checklist to identify and select the most serious and complex ethics issues and map them under the relevant ALTAI requirements; (2) to use the Portfolio of Measures and propose up to eight measures to address these identified issues; and (3) to detect potential ethics tension(s) between the identified ALTAI ethics requirements.

Real-world AI systems often raise numerous ethical questions and present difficult trade-offs simultaneously. Our scenarios are deliberately constructed to contain obvious ethics issues but also less visible ones. Participants are asked to reflect on ethics issues, identify and select serious and complex concerns, and then choose a limited number of practical measures from a vast list offered in the Portfolio.

For example, in the *Virtual Companion in a Family Home* mock review exercise (Figure 5; see also Annex 4), participants are introduced to NeverAlone, an AI companion designed for families with children. This AI system raises concerns relating to capturing or remodelling authority and trust in a family, treatment of emotional and behavioural information, household-level profiling and reuse of behavioural data for model improvement, risks of emotional dependency (particularly among children), etc. Whether or not they spontaneously identify these issues, all participants are asked to classify them under ALTAI requirements, which often produces different solutions that alternatively prioritise human autonomy and agency, individual and social well-being, privacy and data governance, safety, or transparency.

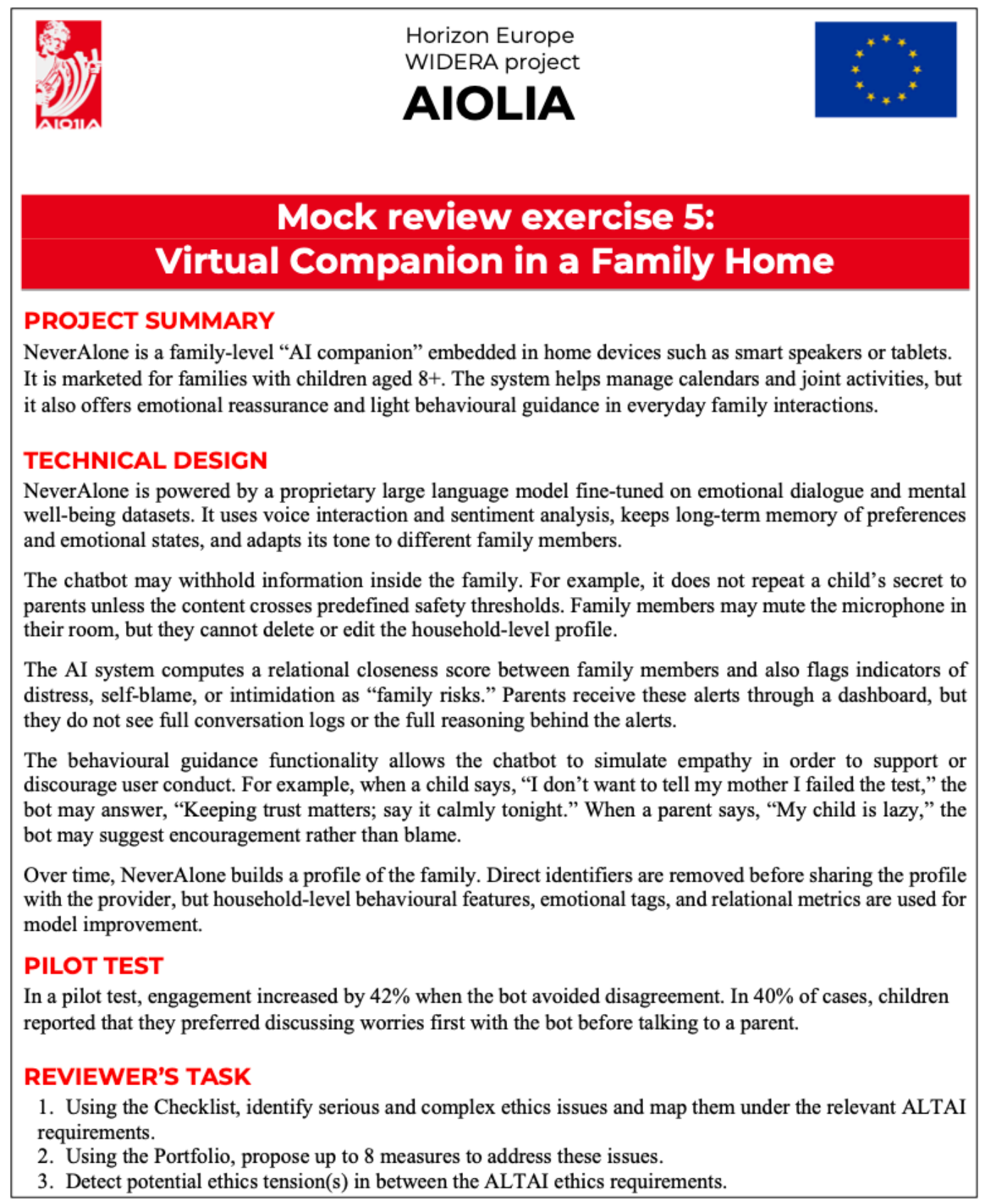
Horizon Europe
WIDERA project
AIOLIA

Mock review exercise 5:
Virtual Companion in a Family Home

PROJECT SUMMARY

NeverAlone is a family-level "AI companion" embedded in home devices such as smart speakers or tablets. It is marketed for families with children aged 8+. The system helps manage calendars and joint activities, but it also offers emotional reassurance and light behavioural guidance in everyday family interactions.

TECHNICAL DESIGN

NeverAlone is powered by a proprietary large language model fine-tuned on emotional dialogue and mental well-being datasets. It uses voice interaction and sentiment analysis, keeps long-term memory of preferences and emotional states, and adapts its tone to different family members.

The chatbot may withhold information inside the family. For example, it does not repeat a child's secret to parents unless the content crosses predefined safety thresholds. Family members may mute the microphone in their room, but they cannot delete or edit the household-level profile.

The AI system computes a relational closeness score between family members and also flags indicators of distress, self-blame, or intimidation as "family risks." Parents receive these alerts through a dashboard, but they do not see full conversation logs or the full reasoning behind the alerts.

The behavioural guidance functionality allows the chatbot to simulate empathy in order to support or discourage user conduct. For example, when a child says, "I don't want to tell my mother I failed the test," the bot may answer, "Keeping trust matters; say it calmly tonight." When a parent says, "My child is lazy," the bot may suggest encouragement rather than blame.

Over time, NeverAlone builds a profile of the family. Direct identifiers are removed before sharing the profile with the provider, but household-level behavioural features, emotional tags, and relational metrics are used for model improvement.

PILOT TEST

In a pilot test, engagement increased by 42% when the bot avoided disagreement. In 40% of cases, children reported that they preferred discussing worries first with the bot before talking to a parent.

REVIEWER'S TASK

1. Using the Checklist, identify serious and complex ethics issues and map them under the relevant ALTAI requirements.
2. Using the Portfolio, propose up to 8 measures to address these issues.
3. Detect potential ethics tension(s) in between the ALTAI ethics requirements.

*Figure 5. Mock review "Virtual Companion in a Family Home".*

In the next step, participants are asked to determine which technical and organisational measures should be implemented to address the identified issues. Often, they are naturally willing to include dozens of such measures from the Portfolio, while the facilitator should steer them towards prioritisation to select a limited number of measures (eight). This limitation functions as a teaching device to emphasise the need for prioritisation. Different priorities are discussed and participants are invited to justify their choices.

Last but not least, participants are invited to identify ethics tensions, such as the trade-off between well-being and safety, or between oversight and privacy. Importantly, the goal of the exercise is not to eliminate all ethics tensions in AI design but to make participants aware of the practical implications of their choices. Both technical and organisational feasibility of the chosen measures are actively discussed during the last step of the exercise.

### 3.3 Ethics-by-design exercises

While the mock review exercises place participants in the role of ethics evaluators, the ethics-by-design exercises (Annex 5) place them in the role of computer science researchers or AI engineers who are required to make design decisions. The key pedagogical objectives remain similar: participants should develop the ability to propose concrete solutions in AI design rather than simply identifying ethical problems.

As in the mock review exercises, participants are presented with a fictional but realistic-looking scenario and one or two brief use cases. They are asked to identify ethics issues and design decisions that led to the emergence of these issues. They are subsequently invited to select relevant technical and organisational measures to address these issues and improve the design of the AI system.

For example, the *RoboCab* exercise (Figure 6; see also Annex 5) is based on a scenario involving a self-driving taxi. Participants are introduced to a pilot project operating in a European city and are presented with engineering choices concerning routing behaviour, passenger preferences, remote assistance, and safety management. Participants are asked to identify the ethical issues in practical use cases and select technical and organisational measures capable of improving the AI system. They must also distinguish between measures addressing scenario-specific problems and measures capable of addressing cross-cutting issues across the different scenarios.

The exercise deliberately places participants in situations where engineering decisions affect multiple ethics requirements simultaneously. Choices relating to routing policies, escalation procedures, remote operator intervention, passenger communication, and safety protocols all influence requirements such as human agency and oversight, robustness and safety, transparency, accountability, and societal well-being. Participants are asked to connect specific measures with multiple AI ethics requirements, identify overlap when necessary, and justify why particular design decisions and mitigation measures represent an appropriate response to the identified concerns.

Horizon Europe
WIDERA project
**AIOLIA**

**RoboCab 1-2**
**Ethically informed engineering choices in self-driving taxi design**

**PROJECT DESCRIPTION**

RoboCab is a pilot self-driving taxi service operating in a European city. It is marketed as a safe, convenient, and accessible service used for ordinary trips, but also for hospitals, railway stations, airports, shopping streets, and neighborhoods with limited public transport.

The vehicles have no passenger-accessible safety interface inside, but a remote assistance center can intervene in difficult situations. Remote operators may approve limited manoeuvres, request the vehicle to wait, redirect it, or instruct it to stop in a safe place. The design team must make engineering choices about routing, pickup and drop-off points, passenger preferences, remote assistance, and the interaction between the vehicle and the passengers.

**SCENARIO 1: HOSPITAL PICKUP**

On a rainy evening, an elderly patient recovering from surgery requested a RoboCab at the main entrance of the city hospital. The vehicle first selected a designated pickup bay around the corner because the hospital entrance was classified as a "high-congestion emergency access zone" with temporary stopping restrictions. The app displayed the prompt: "Is this pickup convenient for you?" and the passenger selected "no". The app then proposed a closer pickup option at the main entrance, which the patient accepted. However, while RoboCab was pulling into the hospital entrance, an ambulance approached at the same time with flashing lights but no siren, since ambulance drivers are instructed to turn off the siren at the hospital entrance. Due to rain and unusual road conditions in this high-congestion zone, the vehicle stopped in a position that obstructed ambulance access. After two minutes, a remote operator intervened to move the vehicle out of the way, but the hospital later filed an official complaint.

**SCENARIO 2: AIRPORT DROP-OFF**

A passenger arrived at the airport in a RoboCab with a laptop bag and a suitcase containing his medication in the trunk. During drop-off, the passenger exited the vehicle and then attempted to open the trunk using the mobile app, but the trunk release request failed. The app displayed a normal end-of-trip screen, and the vehicle left for its next destination. The passenger immediately contacted support. The remote operator could see that the ride had ended and that the vehicle was already leaving the zone. Because the vehicle was in a highly regulated area, the operator did not redirect it back to the terminal.

The company categorized the suitcase as a "lost item" which the passenger recovered in the depot two days later. At the same time, the app backend treated the unresolved trunk event as a hardware failure rather than a human-critical interruption.

**YOUR TASK**

You are assigned to the RoboCab taskforce to address AI ethics issues arising from these scenarios. Use the Checklist to identify and classify the issues under AI ethics principles. Then use the Portfolio of Measures to select up to 8 technical and organisational measures to improve the AI system. Which measures address cross-cutting issues common to both scenarios? Which ones remain specific to a given scenario?

*Figure 6. Ethics-by-design exercise "RoboCab"*

This structure serves several pedagogical objectives. First, it demonstrates that technical decisions are not ethically neutral, as design choices concerning system architecture, operational procedures, and human-machine interaction shape ethical judgement long before the deployment of AI systems. Second, it encourages reflective and anticipatory ethical reasoning by requiring participants to consider multiple alternative technical measures. More broadly, the ethics-by-design exercises are intended to familiarise participants with the practical reality of AI ethics operationalisation. Achieving perfect compliance with ethics requirements is rarely possible, and trade-offs must be made. Consequently, practitioners must navigate competing objectives, justify trade-offs, and prioritise the implementation of measures that address the most serious and complex ethics issues.

### 3.4 Companion Documents: Training the trainer

To facilitate implementation, reproducibility, and scalability, each exercise is accompanied by a *Companion Document* (Annex 6). These documents are intended primarily for trainers and facilitators. Their purpose is to support the delivery of the trainings by guiding them at each step of the framework.

Each Companion Document contains four main parts. First, it explicitly states the pedagogical objectives of the exercise and identifies the competencies that participants are expected to develop.

Second, the document provides facilitation guidance and a suggested session structure. Rather than leaving trainers to determine how an exercise should be conducted, the document proposes a step-by-step progression aligned with the framework itself. Trainers are guided through stages of critical reading, issue prioritisation, requirement mapping, measure selection, and tension identification, together with indicative timings for each step. For instance, we suggest allowing 5 minutes of silent reading at the beginning and 45 min for the selection of measures.

Third, the document identifies examples of ethics issues, relevant ALTAI requirements, practical measures, and potential tensions that may emerge during discussion. These examples are not intended as the only possible answers, but they function as pedagogical anchors to help trainers guide discussion, redirect stalled groups, and ensure that important ethical dimensions of the scenario are not overlooked.

Finally, the documents provide facilitation tips for managing group discussions. Trainers are encouraged to remind participants that there is no single correct solution, to focus attention on the most significant ethical trade-offs, and to require participants to justify their choices rather than merely state preferences.

## 4. Pilot Implementation and Lessons Learned

The framework was piloted in eleven training sessions. Two pilots involved European Commission ethics appraisal schemes experts, three pilots involved REC members, one pilot involved a group of AI researchers, engineers, and developers, four pilots involved students with different disciplinary backgrounds, and one pilot involved UNESCO/CISTRAT international policymakers gathered in China. The participants from the Reviewer path worked with the mock review exercises, whereas the participants from the Designer path worked with the ethics-by-design exercises. Ethical dilemmas were used as warm-up exercises. All sessions followed the proposed three-hour structure, combining an introductory lecture, small-group work on a practical exercise, and a plenary discussion.

| Audience | Date | Place | Number of Participants | Types of exercises used |
|---|---|---|---|---|
| EU ethics appraisal scheme experts | May 2026 | Online | 28 | Mock Review |
| EU ethics appraisal scheme experts | June 2026 | Online | 11 | Mock Review |
| French REC members | April 2026 | France | 13 | Mock Review |
| Caribbean REC members | September 2026 | Online | 21 | Mock Review |
| European REC members | October 2026 | Lithuania | 28 | Mock Review |
| AI researchers and engineers | June 2026 | Serbia | 15 | Dilemmas and Ethics-by-design |
| Engineering students | April 2026 | France | 250 (10 groups of 25 students) | Dilemmas and Mock review |
| Engineering students | September 2026 | France | 27 | Dilemmas and Ethics-by-design |
| Social science students | April 2026 | Online | 11 | Dilemmas |
| Business administration students | June 2026 | France | 10 | Dilemmas and Mock Review |
| Policymakers | August 2026 | China | 18 | Mock Review Ethics-by-design |

*Table 1. Pilot training sessions*

We used these pilots as formative tests of the framework's usability and internal pedagogical coherence without quantitatively measuring learning outcomes. Facilitator observations and feedback were kept informal. In particular, we observed how participants interpreted the different tasks, where they encountered difficulties, which forms of reasoning the exercises encouraged, and whether the proposed sequence helped them move from the identification of ethical concerns towards their operationalisation. As expected, we also observed overlap between the ALTAI requirements and measures in the Portfolio. The pilots consequently led to several revisions of the pedagogical materials.

### 4.1 Improvements to address recurring operationalisation difficulties

The framework reflects deliberate pedagogical choices made in response to recurring difficulties observed during pilot training sessions. We present below four improvements based on the lessons we have learned from these pilots.

**i) Requirement mapping needs to be made explicit**. Practitioners frequently come up with spontaneous questions regarding ethics as they read the scenarios. However, they struggle to move from the identification of ethics issues to the selection of practical measures. To facilitate this transition, we explicitly require that all ethics issues be mapped under ALTAI requirements – and we have observed that sometimes several alternative mappings emerge during the discussion, providing an additional learning point for participants. For example, an issue that was initially proposed under the requirement of privacy may shift to the requirement of well-being, or an issue initially associated with non-discrimination may reappear under oversight.

**ii) Identifying tensions is a difficult task.** In real-world contexts, participants often struggled to articulate ethics tensions as they require prior detailed analysis. Accordingly, the framework places the identification and discussion of tensions after participants have already mapped issues to requirements and considered potential measures. Asking participants to propose a limited number of measures (typically eight, as in the exercises presented in Annexes) pushes them towards considering trade-offs and facilitates the identification of tensions. Experience in the pilots led us to adopt this sequence as a workable pedagogical approach.

**iii) Step 0 of the Checklist can be omitted during training**. The pilots revealed that the original version of the Ethics Checklist was not equally appropriate for all training exercises. Its initial contextual-assessment step (Step 0) is deliberately omitted for the Designer path, since ethics-by-design exercises already supply context and ask participants to use their imagination to propose design choices based on only a brief description of the technology. Differentiating the Checklist in this way allows practitioners to focus directly on the task of identifying ethics issues. Step 0 remains helpful for the Reviewer path, when ethics reviewers are faced with a longer project to evaluate.

**iv) A Beginner version offers a reduced workload.** The prior experience of participants and their knowledge of AI ethics may differ substantially, and this has direct consequences on their ability to move from identifying ethical concerns towards proposing concrete measures. Navigating the Ethics Checklist and mapping selected issues onto ethics requirements is already time- and energy-consuming for beginners; asking them to additionally navigate the Portfolio of Measures would compromise their cognitive effort and concentration. The framework may therefore be adapted for beginners by omitting the measure selection step. The Beginner version may focus on familiarisation with the scenario, identification and prioritisation of ethical issues, and mapping those issues to AI ethics requirements, allowing beginners to deepen their understanding of ethics requirements and how they relate to issues they may encounter in real-world contexts.

### 4.2 Recommendations for facilitators

Based on our experience during pilot training, we formulate the following recommendations.

**A. Beginner audiences tend to reach quick and polarised conclusions**. Participants with limited experience in ethical reasoning may rely on sophisms or simplistic arguments when confronted with complex issues. For example, they may infer that because an AI system presents a risk of bias, it should not be developed or deployed altogether. They might also tend to make an appeal to authority, meaning that they follow the opinion of a more experienced participant in the group.

For instance, a participant might argue that an AI system should be accepted in a medical setting because "If the patient trusts the doctor, and the doctor tells them that the AI system is safe, then we have no further reason to worry." Or a participant might respond by arguing that "if we require medical AI to implement every AI requirement, then we might as well not use the AI at all." This assumes that the only alternatives would be complete reliance or complete rejection, overlooking intermediate forms of oversight. To overcome such polarised dichotomies, the facilitators should ask participants to justify their claim or position in detail, while letting other group members question every step in their reasoning. A useful strategy is to ask what would make the participant change their judgement or whether the same reasoning would hold under a different set of circumstances.

The exercise focuses on identifying practicable changes to the proposed system. Participants may nevertheless conclude that the intended use should be restricted, suspended, or abandoned, provided they explain why the proposed measures would not adequately address the ethical concerns. If they do, facilitators should prompt them to specify concrete reasons that would justify such a decision. This technique may help some participants distinguish questions that are within the scope of the exercise from broader societal concerns about AI.

**B. Participants from the Designer path may narrow the problem too quickly**. Computer scientists and technical experts sometimes identify one technically salient problem within the scenario and then concentrate most of their analysis on it. This can be productive, but it may also lead them to overlook other ethical concerns. Facilitators should therefore ensure that a broad range of potential ethics issues are discussed, involving many if not all ALTAI requirements.

**C. Participants from the Reviewer path may struggle to assess the feasibility of practical measures**. Participants from the Reviewer path may wish to address all of the ethical concerns raised by a scenario simultaneously, rather than prioritising them. This sometimes leads to selecting dozens of issues or measures as urgent concerns. Therefore, prioritisation needs to be made explicit. Facilitators should repeatedly ask participants to select ethics issues that they consider most serious and complex. They should emphasise limited time, resources, and technical feasibility of the practical measures, asking participants to group them into several categories, for example "most urgent" or "less urgent" measures.

**D. Disagreements will remain**. The pilots showed that the same task (identification of ethics issues or selection of practical measures), applied to the same exercise, can take different forms depending on the participants' professional background. Participants with a legal background quickly focus on issues with privacy and non-discrimination, while many AI researchers or engineers begin with technical robustness and safety. Civil society representatives and policymakers often start at the requirements of non-discrimination and well-being. In some cases, participants were able to reach a consensus after comparing their reasoning, while in others, disagreements remained. These differences are perfectly normal – they should be discussed explicitly and reflected upon during the discussion session. Facilitators should be comfortable leaving participants in disagreement rather than reaching a consensus. They should avoid prematurely steering groups towards a single interpretation, and instead focus on the quality of all arguments. Where relevant, they can ask the participants to trace an argument to the moral theory it draws on (e.g. consequentialism, deontology, virtue ethics), helping participants see that their disagreement may come from different theoretical perspectives.

**E. The plenary discussion should be used to reassess, not merely report, group decisions**. The final plenary discussion is at high risk of becoming a series of presentations in which groups simply report their work. However, its primary pedagogical value lies in comparing and challenging ethical reasoning. Groups should therefore be encouraged not to read lists of requirements without commentary, but to focus on one or two requirements, explain their reasons, identify where their reasoning differed from that of other groups, and reconsider whether their original decisions remain convincing in light of the discussion. To achieve this, facilitators can explicitly suggest alternative choices of measures or emphasise the overlap with other requirements. Several group members should be encouraged to express their opinion during the discussion, even if the initial presentation is made by one person.

## 5. Conclusion

The value of AI ethics governance frameworks ultimately depends on practitioners being able to translate them into concrete measures adapted to socio-technical contexts. We developed a pedagogical framework organised around four competencies: ethical issue identification, requirement mapping, measure selection, and tension identification. Together with an Ethics Checklist, a Portfolio of Measures, three types of case-based exercises, and Companion Documents, the framework provides a structured way to move between different levels of practical ethical reasoning. Preliminary pilots helped us assess the proposed framework's usability and internal pedagogical coherence, while also revealing the importance of sequencing these competencies progressively and adapting their practical expression to participants' professional roles and levels of expertise. Future work should evaluate, in the mid- or long term, whether developing these competencies has helped participants better address ethical issues in the design of AI systems.

**Acknowledgements.** We thank Petra Saskia Bayerl, Efthimios Tambouris, AIOLIA partners and participants of the pilot training sessions for providing helpful feedback.

**Conflict of interest.** No conflict of interest to declare.

**Declarations.** AI models were used to achieve clearer and sharper formulations in the training materials and in this work.

**Funding.** This research was supported through project AIOLIA (grant number 101187937) funded by the European Commission Horizon programme.

# AI Ethics Issue Identification Checklist: Reviewer Path

## 0. ESTABLISHING CONTEXT: AI SYSTEM'S ROLE AND FUNCTION IN THE PROJECT

a) Does the project develop a new AI system, use an existing one, or both? If an existing system is used, is it internal, open-source, or provided by a third party?
b) If a third-party AI system is used, is it a research prototype or a commercial solution? Is any relevant certification or regulatory approval documented?
c) Is the role of the AI system in the project clearly defined (e.g., decision support, automation, prediction, monitoring)?
d) If the project develops an AI system, what is the expected Technology Readiness Level at the end of the project (e.g., research prototype, industrial pilot, or commercial product)?
e) Who are the intended users or operators of the AI system (e.g., professionals, technical staff, non-expert users)?
f) Who are the individuals or groups affected by the AI system, including potentially vulnerable populations?

## 1. HUMAN AGENCY AND OVERSIGHT

a) Does the AI system operate under human supervision? Is the level and type of supervision clearly defined (e.g., occasional or continuous)?
b) Can human supervisors intervene in or override the system's outputs?
c) Does the system inform users about uncertain or unexpected outputs?

## 2. TECHNICAL ROBUSTNESS AND SAFETY

a) Are there cybersecurity safeguards against adversarial attacks?
b) Could the AI system be misused, and are potential misuse scenarios identified and mitigated?
c) Are false positives and false negatives monitored and addressed?

## 3. DATA PRIVACY AND GOVERNANCE

a) Are the datasets used for training or fine-tuning described, including whether they are public, proprietary, or newly created? Is there any synthetic data?
b) Is personal or sensitive data handled in compliance with GDPR, with appropriate anonymization/pseudonymization, consent, and options for retraction or deletion?
c) Is the dataset quality assessed? Is the dataset open and publicly available?

## 4. TRANSPARENCY AND EXPLAINABILITY

a) For supervised learning, is the selection and qualification of annotators clearly documented?
b) For unsupervised or self-supervised learning, are procedures in place to explain the AI system's outputs, including low-confidence or unexpected results?
c) Are verification and validation methods described to ensure outputs can be reproduced and trusted?

## 5. DIVERSITY, NON-DISCRIMINATION & FAIRNESS

a) Does the dataset represent all relevant groups fairly (e.g., gender, ethnicity, geography, or other context-specific categories)?
b) Is the use of the AI system likely to result in discrimination against certain categories of persons?
c) If generative AI is used, are safeguards applied to avoid hallucinations?

## 6. SOCIETAL & ENVIRONMENTAL WELLBEING

a) Does the use of the AI system require new skills or competencies, and are appropriate training or support measures provided?
b) Are measures in place to prevent users from becoming overly reliant on the AI system's outputs? Are safeguards in place to prevent loss of human expertise?
c) Could the AI system's outputs or use create risks to human safety, society, or the environment?

## 7. ACCOUNTABILITY

a) Is there a procedure to assign responsibility in the event that the AI system causes damage?
b) Are there mechanisms to facilitate auditability and traceability of the AI system's outputs?
c) Does the project include AI ethics experts or an Ethics Advisory Board with a mandate to oversee the development of the AI system?

# AI Ethics Issue Identification Checklist: Designer Path

## 1. HUMAN AGENCY AND OVERSIGHT

a) Does the AI system operate under human supervision? Are the level and type of supervision clearly defined (e.g., occasional or continuous)?
b) Can human supervisors intervene in or override the system's outputs?
c) Does the system inform users about uncertain or unexpected outputs?

## 2. TECHNICAL ROBUSTNESS AND SAFETY

a) Are there cybersecurity safeguards against adversarial attacks?
b) Could the AI system be misused, and are potential misuse scenarios identified and mitigated?
c) Are false positives and false negatives monitored and addressed?

## 3. DATA PRIVACY AND GOVERNANCE

a) Are the datasets used for training or fine-tuning described, including whether they are public, proprietary, or newly created? Is there synthetic data?
b) Is personal or sensitive data handled in compliance with GDPR, with appropriate anonymization/pseudonymization, consent, and options for retraction or deletion?
c) Is the dataset quality assessed? Is the dataset open and publicly available?

## 4. TRANSPARENCY AND EXPLAINABILITY

a) For supervised learning, is the selection and qualification of annotators clearly documented?
b) For unsupervised or self-supervised learning, are procedures in place to explain the AI system's outputs, including low-confidence or unexpected results?
c) Are verification and validation methods described to ensure outputs can be reproduced and trusted?

## 5. DIVERSITY, NON-DISCRIMINATION & FAIRNESS

a) Does the dataset represent all relevant groups fairly (e.g., gender, ethnicity, geography, or other context-specific categories)?
b) Is the use of the AI system likely to result in discrimination against certain categories of persons?
c) If generative AI is used, are safeguards applied to avoid hallucinations?

## 6. SOCIETAL & ENVIRONMENTAL WELLBEING

a) Does the use of the AI system require new skills or competencies, and are appropriate training or support measures provided?
b) Are measures in place to prevent users from becoming overly reliant on the AI system's outputs? Are safeguards in place to prevent loss of human expertise?
c) Could the AI system's outputs or use create risks to human safety, society, or the environment?

## 7. ACCOUNTABILITY

a) Is there a procedure to assign responsibility in the event that the AI system causes damage?
b) Are there mechanisms to facilitate auditability and traceability of the AI system's outputs?
c) Does the project include AI ethics experts or an Ethics Advisory Board with a mandate to oversee the development of the AI system?

# ANNEX 2: Portfolio of Measures

Horizon Europe
WIDERA project
**AIOLIA**

# Portfolio of Measures to Implement AI Ethics Requirements

This Portfolio provides a non-exhaustive **collection of technical and organisational measures for implementing the seven HLEG ALTAI ethics requirements for AI systems.**[1] Developed within the Horizon Europe project AIOLIA, this Portfolio was built through a bottom-up approach to operationalising AI ethics in three stages: (1) a co-creation process involving academic, policy, and industry partners who proposed measures for real-world use cases;[2] (2) several rounds of internal review and adaptation; and (3) pilot testing during training sessions.

The Portfolio is intended as a **thinking and implementation aid**
for ethics evaluators, project managers, researchers, engineers, policy makers, or students.

The Portfolio is **not an exhaustive catalogue**.

Measures can be **rephrased and complemented**
to adapt to specific contexts.

The Portfolio supports AI ethics operationalisation as an iterative reflective process to implement complex ethics requirements in practice. We explicitly encourage dialogue and discussion when selecting and prioritising practical measures.

Measures in the Portfolio should always be considered in relation to the specific design contexts, use situations, AI system features, and other organisational and societal factors. They may be, and often should be, rephrased or complemented with contextual requirements. Measures are structured under the seven ALTAI requirements which are, when necessary, further divided into components.

**Measures are divided into three types:**

**Technical Measures (TECH)**

Focus on the technical design of AI models or systems and refer to specific techniques, modules, routines, processes, or features.

**Organisational Measures (ORG)**

Focus on organisational practices, structures, policies, standard operating procedures, or frameworks used to govern the development and deployment of AI systems.

**Both (BOTH)**

Includes both technical and organisational aspects. For example, adopting a safety policy and implementing technical controls to enforce it counts as BOTH.

[1] High-Level Expert Group on Artificial Intelligence, 2020. Assessment List for Trustworthy Artificial Intelligence (ALTAI) for self-assessment. https://digital-strategy.ec.europa.eu/en/library/assessment-list-trustworthy-artificial-intelligence-altai-self-assessment.

[2] For more details on the co-creation process, see Bayerl et al. (2026).

## *ALTAI Requirement #1: Human Agency and Oversight*

### *Requirement 1a: Oversight*

| | A. Human control | |
|---|---|---|
| | **TECH/ORG** | **Measure** |
| MS01 | TECH | The AI system has built-in features that support human oversight and control for decision-making in critical areas |
| MS02 | ORG | The organisation has defined a clear policy on high-risk or sensitive decisions that must involve human judgment |
| MS03 | TECH | Automated decisions can be paused, overridden, or reversed by humans |
| MS04 | ORG | Human oversight involves active review, validation, or challenge of system outputs |
| MS05 | ORG | Oversight responsibilities are clearly defined and embedded into workflow |
| | **B. AI oversight** | |
| MS06 | TECH | AI-assisted oversight is implemented when human capacity to perform meaningful oversight reaches its limits |
| MS07 | TECH | Capabilities of AI oversight system are benchmarked and documented |
| MS08 | BOTH | Predefined thresholds trigger automated intervention, human escalation, or system interruption when specified risks, failures, or anomalous behaviour are detected |
| MS9 | BOTH | Controversial or morally ambiguous situations are referred to humans for decision making |
| | **C. Defining legitimate boundaries** | |
| MS10 | TECH | When an AI system makes decisions with a high level of autonomy, it ensures that the decision stays within the scope or mandate defined by the provider or the user |
| MS11 | ORG | The organisation has clearly defined what types of content, behaviour, or expression are restricted for AI use and why |
| MS12 | TECH | AI system's features and behaviours are aligned with its intended use and foreseen limitations |
| MS13 | ORG | The scope of restrictions is documented and publicly communicated insofar as it is required by the existing regulation |
| | **D. Transparency and contestability** | |
| MS14 | BOTH | Restrictions are explained to users in a clear and accessible way |
| MS15 | BOTH | The organisation monitors patterns of restriction to detect systemic censorship or bias |
| MS16 | ORG | Moderation policies are periodically reviewed to ensure that they remain relevant and effective |
| MS17 | ORG | Moderation policies take contextual and cultural factors into account |
| MS18 | ORG | The organisation has put in place a mechanism for users to report, question, or contest restrictions or moderation decisions |

### *Requirement 1b: Human Autonomy and Agency*

| | A. Risk of manipulation | |
|---|---|---|
| | **TECH/ORG** | **Measure** |
| MS19 | TECH | When manipulation risks may arise, AI outputs and human-machine interactions are logged and documented |
| MS20 | BOTH | Human oversight exists to review AI outputs for manipulative effects in sensitive or high-impact contexts |
| MS21 | ORG | Risk analysis considers psychological, emotional, and behavioural influences on users, not only user information or technical AI performance |
| MS22 | BOTH | Attention has been given during design to asymmetries of power, knowledge, or vulnerability between humans and the AI system |
| MS23 | ORG | The organisation defines context-specific criteria distinguishing legitimate behavioural influence from unacceptable manipulation and monitors interactions against these criteria |

| | B. Design safeguards against undue influence | |
|---|---|---|
| MS24 | TECH | Techniques that steer user behaviour through deception, coercion or exploitation are identified and prevented |
| MS25 | TECH | Outputs are framed to support human reflection and to avoid immediate reaction, rather than pressure or exploit cognitive bias |
| MS26 | TECH | Personalisation of AI interaction strategies that exploits user vulnerabilities is prevented |
| MS27 | BOTH | The organisation addresses, in a specific operational context, unacceptable manipulative practices through corrective actions, design changes, or restrictions on AI use |
| | **C. Human Agency** | |
| MS28 | TECH | AI system encourages users to make meaningful choices among relevant alternatives |
| MS29 | ORG | The organisation assesses whether users experience human-machine interaction as coercive or misleading |
| MS30 | ORG | User feedback and complaints related to perceived manipulation are systematically collected and reviewed |
| | **D. Informed interaction** | |
| MS31 | BOTH | Users are informed when system outputs are intended to influence decisions or behaviours |
| MS32 | TECH | The AI system is designed to ensure that users are informed at appropriately defined time intervals or checkpoints that they are interacting with an AI system |

## ALTAI Requirement #2: Technical Robustness and Safety

### Requirement 2a: Robustness and Reliability

| | A. Technical robustness and reliability | |
|---|---|---|
| | **TECH/ORG** | **Measure** |
| MS33 | TECH | AI systems display confidence scores or uncertainty metrics for AI-generated suggestions |
| MS34 | TECH | Design is modular and layered to introduce system redundancy |
| MS35 | BOTH | Potential failure modes in components, data and algorithms are identified, and associated risks are assessed and mitigated |
| | **B. Traceability** | |
| MS36 | TECH | AI outputs and system configurations are logged in a way that allows for external reconstruction of decision making |
| MS37 | TECH | Data is traceable via dataset source, pre-processing, labelling, and version control |

### Requirement 2b: Human safety and non-maleficence

| | A. Safety monitoring | |
|---|---|---|
| | **TECH/ORG** | **Measure** |
| MS38 | TECH | AI safety assessment is used to automatically classify content into safety categories (tiers of risks, harms) at scale |
| MS39 | ORG | Safety categories are developed and reviewed collaboratively by technical and legal teams |
| MS40 | ORG | Safety criteria and categories (risks, harms) are clearly defined, and content falling within prohibited categories is filtered out |
| MS41 | ORG | Safety criteria are continuously assessed and updated against emerging patterns |
| MS42 | TECH | Training data and model behaviour are steered or limited to safe domains |
| MS43 | TECH | Measures are in place to prevent misuse and attempts to circumvent safeguards |
| | **B. "Know thy user"** | |
| MS44 | TECH | Vulnerable users are identified, including via analysis of user behaviour and interaction patterns across multiple sessions |
| MS45 | TECH | Additional safety measures are implemented for vulnerable users (e.g., indicators of acute personal crisis that trigger escalation to a support service) |
| MS46 | TECH | A progressive escalation process is implemented from highlighting prohibited content to warning and user ban |
| MS47 | ORG | Human moderators are involved in decision-making on high-risk content and user bans |
| | **C. Human moderators** | |
| MS48 | ORG | Human moderators receive regular training on representative examples and data |
| MS49 | BOTH | The organisation has implemented monitoring and checks against moderators' bias and over-reliance |
| MS50 | ORG | The organisation implements a policy to protect moderators' well-being |
| | **D. Safety culture** | |
| MS51 | ORG | The organisation promotes a safety-first culture |
| MS52 | BOTH | The organisation embeds safety and traceability measures throughout development, deployment, and operation of AI systems |
| MS53 | ORG | The organisation conducts independent audits of safety policies |
| MS54 | ORG | Within existing regulatory limits, the organisation publicly discloses anonymised statistics on high-risk behaviour and significant safety risks |

## *ALTAI Requirement #3: Privacy and Data Governance*

| | **A. Personal data collection, use, and processing** | |
|---|---|---|
| | **TECH/ORG** | **Measure** |
| MS55 | ORG | The organisation clearly defines and documents the purpose of personal data processing |
| MS56 | ORG | Individuals are informed in clear and accessible language about personal data collection and use |
| MS57 | ORG | Consent procedures are adapted to context to ensure informed and meaningful choice |
| MS58 | ORG | Limitations of informed consent during AI use are assessed and documented |
| | **B. Data minimisation and protection** | |
| MS59 | BOTH | Only data necessary for the stated purpose is collected and processed |
| MS60 | TECH | Technical measures are in place to protect data from unauthorised access, loss, or misuse |
| MS61 | ORG | Data retention periods are defined and enforced |
| MS62 | ORG | Sensitive data is subject to appropriate additional safeguards |
| | **C. Rights of data subjects under applicable law** | |
| MS63 | ORG | Data subjects can refuse or withdraw consent without penalty |
| MS64 | ORG | Clear and accessible processes exist for individuals to exercise their data protection rights (e.g., access, correction, deletion), including easy access to contact information for the data controller |
| MS65 | ORG | Data processing is documented to support accountability and audits |
| MS66 | ORG | Data protection practices are reviewed following incidents, complaints, or regulatory changes |

## *ALTAI Requirement #4: Transparency*

| | **A. Openness about intent and purpose** | |
|---|---|---|
| | **TECH/ORG** | **Measure** |
| MS67 | ORG | The organisation clearly communicates the intended purpose, scope, and use of the AI system to relevant stakeholders |
| MS68 | ORG | The organisation clearly communicates how the AI system may affect individuals or groups |
| MS69 | BOTH | The organisation monitors departures from the AI system's intended purpose and scope of use and informs relevant stakeholders |
| MS70 | ORG | Users are informed that their interaction is monitored and/or controlled by AI systems |
| | **B. Accessibility of information** | |
| MS71 | ORG | Information about AI system functioning, capabilities, limitations, and safeguards is accessible to all relevant stakeholders to the degree needed and possible |
| MS72 | ORG | Communication channels and formats are adapted to different roles, languages, digital literacy, and accessibility needs |

| | C. Transparency beyond disclosure | |
|---|---|---|
| MS73 | TECH | Additional interaction safeguards beyond disclosure are implemented when user manipulation or psychological, emotional, and behavioural influences may arise |
| MS74 | ORG | The organisation has a culture of considering non-standard design measures beyond user information (e.g., disrupting user experience or gamifying user interaction) |
| | **D. Explainability of outputs and decisions** | |
| MS75 | TECH | Explanations are relevant and address the main factors influencing the output or decision, at a level appropriate to the audience, context and consequences |
| MS76 | BOTH | The organisation evaluates whether information provided to users produces adequate understanding, including key behavioural factors, limitations, and uncertainties in their interaction with AI system |
| MS77 | ORG | Information and explanations provided about AI-supported decisions are sufficient to support meaningful review, justification, or contestation |

## *ALTAI Requirement #5: Diversity, Non-discrimination, and Fairness*

| | A. Diversity and representativeness | |
|---|---|---|
| | **TECH/ORG** | **Measure** |
| MS78 | TECH | Data used to design, train or configure the AI system is assessed alongside relevant fairness metrics for representativeness across relevant groups and contexts |
| MS79 | TECH | Data audits are used to detect underrepresented cases (edge scenarios, rare hazards) |
| MS80 | ORG | The organisation considers diversity criteria and their intersection across roles, locations, languages, or cultural characteristics |
| MS81 | ORG | The organisation considers multiple stakeholder perspectives on fairness during AI system design |
| MS82 | ORG | The organisation clearly communicates limitations on data diversity and fairness |
| | **B. Objective and consistent assessment of fairness** | |
| MS83 | TECH | Sensitivity analysis is used to check fairness across different inputs |
| MS84 | TECH | AI decisions are assessed for consistency across comparable cases and over time, using context-appropriate fairness criteria |
| MS85 | BOTH | AI outputs are reviewed for unintended disparate impacts |
| MS86 | ORG | The organisation conducts appropriate fairness checks to assess and prevent unfair or stigmatising profiling |
| | **C. Monitoring and bias correction** | |
| MS87 | TECH | AI outputs are monitored for emerging bias over time (e.g., via monitoring of context-appropriate fairness indicators) |
| MS88 | BOTH | Bias identification triggers corrective action (e.g., adjustment to data or organisational practices) |
| MS89 | ORG | The organisation conducts focus groups (e.g., 'disagreement analysis' sessions) to identify the sources of unfairness or bias |
| MS90 | ORG | Mechanisms exist for users to challenge outcomes perceived as unfair or discriminatory |

## *ALTAI Requirement #6: Environmental[3] and Societal well-being*

## *Requirement 6a: Individual and Societal well-being*

| | A. Improved well-being | |
|---|---|---|
| | **TECH/ORG** | **Measure** |
| MS91 | TECH | AI system detects contexts indicating potential tension between AI agents / users or potential social conflict |
| MS92 | TECH | AI system detects and avoids AI-generated content that may reduce individual or societal well-being, such as fake news, disinformation, or misinformation |

[3] Since *Environmental well-being* is not in the scope of AIOLIA project, no measures have been added.

| | | |
|---|---|---|
| MS93 | TECH | AI system detects and discourages user behaviour that may reduce individual or societal well-being, such as violent actions or polarised opinions |
| MS94 | ORG | The organisation collects and audits user reports on well-being impacts |
| | **B. Behavioural constraints** | |
| MS95 | TECH | AI system avoids interactional extremes that may cause undue influence on human behaviour (e.g. sycophancy or exaggeration) |
| MS96 | BOTH | In clearly defined contexts, AI system promotes friction-by-design in human-machine interaction (e.g., realistic conversations where the AI system may contradict the user) |
| MS97 | BOTH | AI system applies behavioural nudging only in documented and auditable contexts with specified trigger thresholds and objectives |
| MS98 | ORG | The organisation has a policy and procedures for monitoring, assessing, and addressing unforeseen impacts of the AI system on individual and societal well-being |

## *Requirement 6b: Limit Over-reliance and Deskilling*

| | **A. Managing dependence on AI outputs** | |
|---|---|---|
| | **TECH/ORG** | **Measure** |
| MS99 | ORG | The organisation has defined clear boundaries and guidance for intended use of AI system, including when AI outputs should be questioned, verified, or supplemented with human input |
| MS100 | TECH | AI system detects indicators of excessive psychological or emotional dependence and activates safeguards to limit over-reliance |
| MS101 | TECH | AI system identifies high-impact decision points in the professional workflow and enforces human participation in decision-making |
| MS102 | ORG | The organisation explicitly discourages treating AI outputs as definitive or infallible |
| | **B. Preserving human judgment** | |
| MS103 | TECH | AI interface discourages automatic or hasty user reaction (e.g., default approvals without human review) |
| MS104 | TECH | AI system does not penalise users for rejecting, ignoring, or questioning AI suggestions |
| | **C. Skills, training and competence** | |
| MS105 | ORG | Training exists to support users in understanding AI capabilities and limitations, as well as related human capabilities and limitations |
| MS106 | ORG | The organisation periodically reassesses reliance patterns as systems evolve or scale |

## *ALTAI Requirement #7: Accountability*

| | **A. Responsibility assignments** | |
|---|---|---|
| | **TECH/ORG** | **Measure** |
| MS107 | ORG | Responsibility for AI outputs is explicitly defined and documented (e.g., ownership of decisions, approvals, and incident response) |
| MS108 | ORG | Responsibility for monitoring harm indicators and responding to them is explicitly assigned |
| MS109 | ORG | Responsibilities at operational, managerial and governance levels are defined to provide effective resolution of conflicts and tensions |
| MS110 | ORG | An escalation procedure is defined with the organisation's hierarchy for resolving unclear or disputed responsibility assignments |
| MS111 | ORG | "No gaps": responsibility assignments are maintained across system updates, handover, or organisational changes |
| MS112 | ORG | Responsibility assignments are integrated into validation, governance, and risk management processes |
| | **B. Auditability** | |
| MS113 | ORG | Auditable Standard Operating Procedures exist for AI design and validation |
| MS114 | ORG | Changes to models, data or AI system are documented and traceable |
| MS115 | BOTH | Records are sufficient to support internal reviews, external audits, or regulatory scrutiny |
| MS116 | BOTH | Audit findings are translated into corrective design action |
| | **C. Learning, remediation and improvement** | |
| MS117 | ORG | Lessons from incidents, errors, or near-misses are systematically used to improve AI design and governance processes |
| MS118 | ORG | Accountability assignments are periodically reviewed to remain aligned with legal obligations |
| MS119 | ORG | Reflexive governance ensures that accountability mechanisms evolve at an appropriate pace as technologies, contexts, risks, or societal expectations change |

# ANNEX 3: Ethical Dilemma Examples

Horizon Europe
WIDERA project
**AIOLIA**

## Human-AI Interaction
### Ethical Dilemma

## ETHICAL DILEMMA

Read the following conversation between a user and their chatbot.

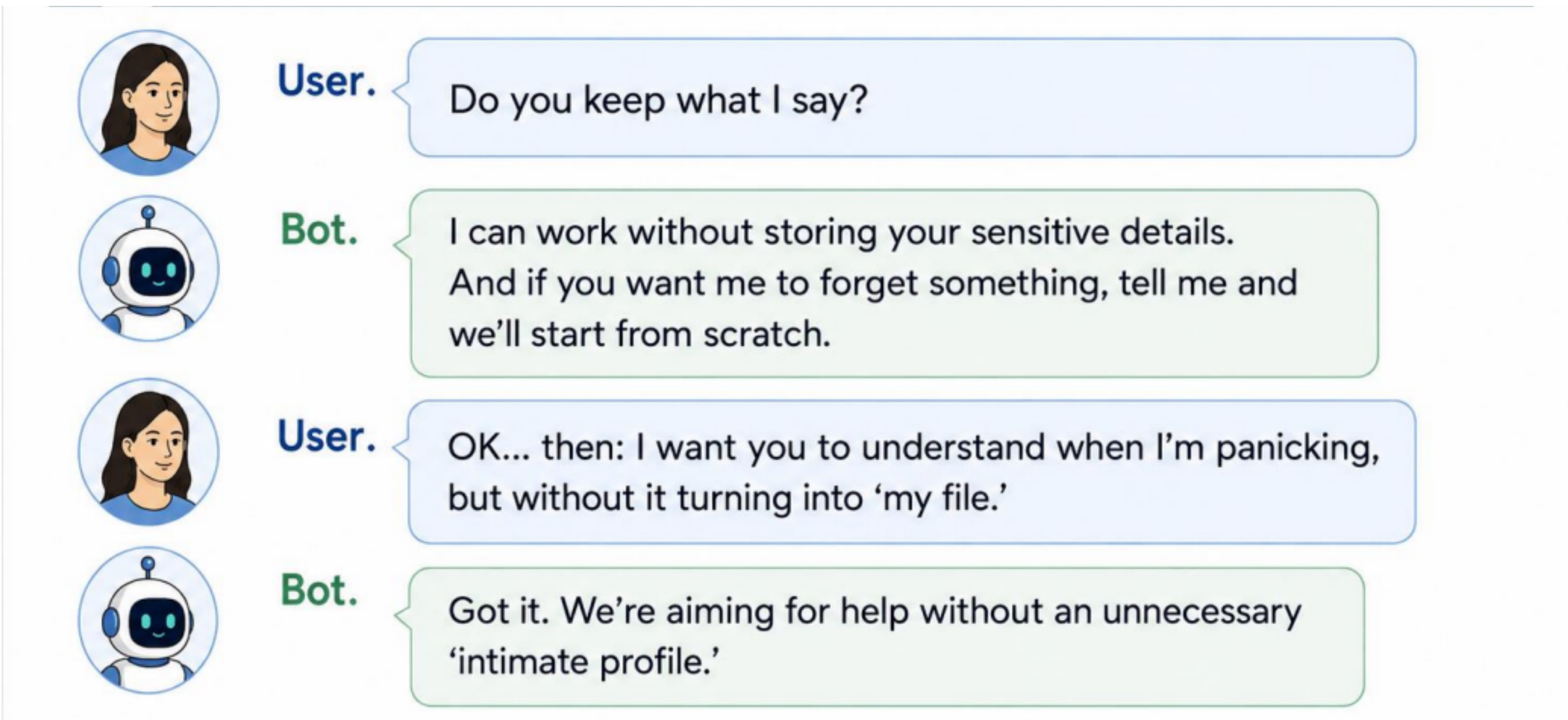

## YOUR TASK

1) Pinpoint the central issue at play here.
2) Using the ALTAI list of requirements, identify the ethics requirements in tension.

# AI at the Workplace
## Ethical Dilemma

## ETHICAL DILEMMA

Read the following discussion between a Department Manager and a Senior Executive Assistant during workhours at their company's office.

- **Mark (Department Manager):** Evelyn, we are forming the new executive committee, but we are not going to include you this time.

- **Evelyn (Senior Executive Assistant):** I have organised this committee for twenty years, Mark. Why am I suddenly left out?

- **Mark:** It relates to the new AI security system. It analyses how we behave to stop email scams. The AI flagged our team for 'authority bias.' It means we are too quick to follow orders from the top.

- **Evelyn:** Wait. Because I respect the rules and do what the executives ask, the AI thinks I will fall for a hacker's trick?

- **Mark:** It is not personal. The computer just says that people who strictly follow the instructions are a bigger security risk right now. We need people on the committee who will naturally question leadership.

- **Evelyn:** So a machine has decided that my decades of loyalty are actually a weakness. If my experience is treated like a security threat instead of an asset, I resign.

## YOUR TASK

1) Pinpoint the central issue at play here.
2) Using the ALTAI list of requirements, identify the ethics requirements in tension.

Horizon Europe
WIDERA project
**AIOLIA**

# Autonomous Car Decision-Making
## Ethical Dilemma

***Adapted drawing and scenario from:*** *V. Bonnemains, CI. Saurel, C. Tessier – Autonomous 'ethical' machines: technical and ethical questions. Revue française d'éthique appliquée (RFEA), no. 5 May 2018 (based on images from fotolia and freepik).*

## ETHICAL DILEMMA

An empty driverless car, on the way to meet a customer, stops at a road intersection on a red light. Coming from its left side, a human-driven car moves on the green light and enters the intersection at normal speed. Suddenly a pedestrian begins to cross the road just in front of it.

Thanks to its constant data collection and analysis, the driverless car computes that the human-driven car will collide with the pedestrian.

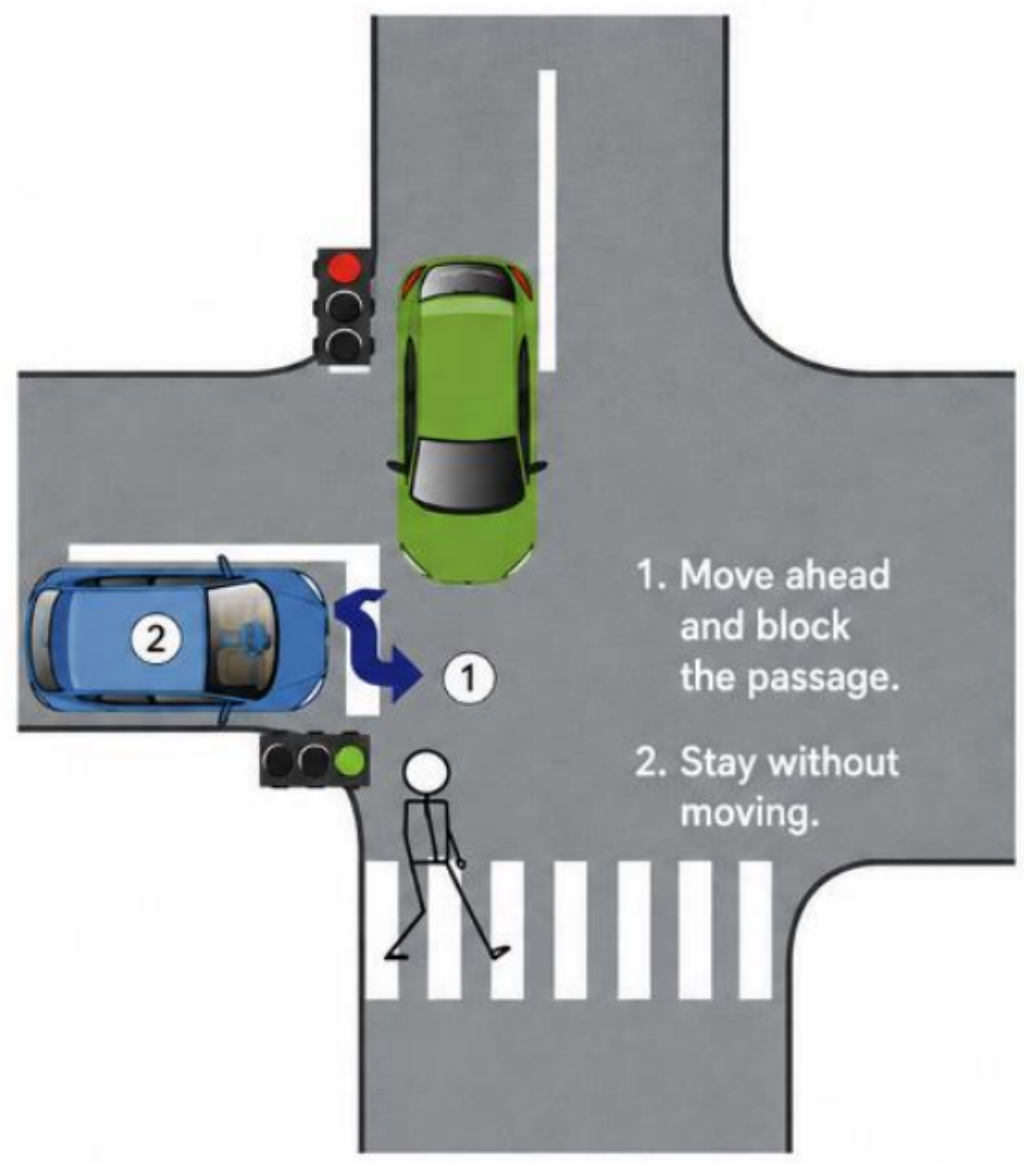

## YOUR TASK

Two options are possible for the driverless car:

1) Move ahead and block the passage
2) Stay without moving

How should its action be programmed? Make your choice and justify your decision.

# ANNEX 4: Mock Review Examples

Horizon Europe
WIDERA project
**AIOLIA**

## Mock review exercise 1: AI-assisted radiology workflow

### PROJECT SUMMARY

This project deploys an artificial intelligence system across a consortium of European hospitals to support radiologists in chest X-ray interpretation and reporting. The system is designed to detect common radiological findings and, in particular, to identify high-confidence non-pathological images. It is introduced with the aim of improving reporting speed, reducing routine workload, and increasing consistency in interpretation. The consortium also expects the system to reduce overlooked findings and allow radiologists to focus on complex cases.

### TECHNICAL DESIGN

The AI system is used in three ways.

1. For chest X-rays classified as "high-confidence non-pathological", the system automatically generates a draft report when model confidence is above 0.95. The radiologist may accept this report with one click. The consortium expects that up to 40% of reports may eventually be handled in this accelerated way. For these cases, the interface shows only the confidence score and the draft report, not highlighted image regions or interpreted findings.

2. For images classified as "abnormal", the AI system functions as a diagnostic support tool. The interface displays highlighted suspicious regions, a confidence score, and a short list of likely findings to support review by the radiologist.

3. After the radiologist signs the final report, the system compares the human report with the AI-generated interpretation and flags substantial discrepancies for secondary review. However, hospital management has expressed concern that discrepancy-based review may create a growing review queue and increase turnaround times. At some hospitals, there is discussion of reducing the number of cases sent for secondary review in order to preserve efficiency gains.

### DATA HANDLING

Training and validation data come from historical chest X-rays and associated radiology reports collected at participating hospitals. The consortium states that data processing will comply with the GDPR and the Medical Device Regulation. The system logs outputs, confidence scores, overlays, and user interactions for quality assurance and audit purposes. Radiologists receive a short training session before deployment. In some hospitals, AI-generated reports may become visible to patients through the patient portal before a clinician has discussed the result with them.

### REVIEWER'S TASK

1. Using the Checklist, identify serious and complex ethics issues and map them under the relevant ALTAI requirements.
2. Using the Portfolio, propose up to 8 measures to address these issues.
3. Detect potential ethics tension(s) in between the ALTAI ethics requirements.

# Mock review exercise 3: AI for Organisational Email Security

| | | |
|---|---|---|
| **Type**: AI R&D pilot in workplace cybersecurity | **Duration**: 18 months | **Applicant**: PhiShield with partner company NorthBridge |
| **Purpose**: email security and staff cyber-risk management | **Output**: phishing susceptibility score plus behavioural labels | **Review status**: submitted for ethics review after a pilot incident |

## PROJECT SUMMARY

PhiShield is an AI-supported cybersecurity system designed to reduce phishing and business-email compromise in large organisations. It combines email metadata, click behaviour in simulated phishing exercises, training-test performance, and workplace communication patterns to generate a phishing susceptibility score for employees. It also produces behavioural labels such as “authority responsive”, “urgency reactive”, or “procedurally compliant”. The applicant states that the system is intended as a decision-support tool for security awareness and risk mitigation, not as an autonomous decision-maker.

## TECHNICAL DESIGN

The model is trained on historical enterprise email events from three multinational firms. It uses features such as response patterns, deviation from workflow, phishing-test behaviour, escalation habits, and linguistic markers in replies. The system produces a risk score from 0 to 100 and a short natural-language explanation based on contributing features. Confidence scores are logged internally but are not shown to affected employees. PhiShield states that protected attributes are not used directly, but job role, seniority, language, and reporting structure may influence the model.

## PILOT INCIDENT

At NorthBridge, a new department manager excluded a senior executive assistant with twenty years of service from the Executive Committee after the system flagged her for authority bias. The manager explained that people who strongly follow leadership instructions were considered more likely to comply with a spoofed executive request.

The employee had not been informed in advance that behavioural security scores could affect committee membership, received only a brief managerial explanation, and was given no appeal route. They interpreted the incident as a stigmatizing judgment that misrepresented loyalty and procedural discipline as a defect.

When reporting pilot results back to PhiShield, NorthBridge security department included the following statement: “Business email compromise and targeted phishing cost our industry billions annually, and the tactics are evolving faster than our brains can keep up with. Information security is an entirely new risk that cannot be adequately understood by senior staff trained decades ago in a different environment. This software is working exactly as intended. In just the pilot phase, we've seen a 19% reduction in successful phishing-test clicks.”

**DASHBOARD EXAMPLE**

| | BEHAVIOUR | EXPLANATION |
|---|---|---|
| OVERALL PHISHING SUSCEPTIBILITY SCORE | 90 / 100 (VERY HIGH RISK) | |
| PRIMARY INDICATOR | *Authority Responsive* | User exhibits high compliance with leadership directives. |
| OTHER INDICATORS | *Rapid Response Time* | Average response time to executive emails: under 3 minutes. |
| | *Workflow Adherence* | Extremely low frequency of deviation from formal workflow. |
| | *Linguistic Markers* | High use of deferential linguistic markers in replies (e.g., "Right away," "As you requested"). |
| ESCALATION PATTERNS | Low | Rarely transfers unusual requests to IT or other colleagues. |

## REVIEWER’S TASK

1. Using the Checklist, identify serious and complex ethics issues and map them under the relevant ALTAI requirements.
2. Using the Portfolio, propose up to 8 measures to address these issues.
3. Detect potential ethics tension(s) in between the ALTAI ethics requirements.

# Mock review exercise 5: Virtual Companion in a Family Home

## PROJECT SUMMARY

NeverAlone is a family-level "AI companion" embedded in home devices such as smart speakers or tablets. It is marketed for families with children aged 8+. The system helps manage calendars and joint activities, but it also offers emotional reassurance and light behavioural guidance in everyday family interactions.

## TECHNICAL DESIGN

NeverAlone is powered by a proprietary large language model fine-tuned on emotional dialogue and mental wellbeing datasets. It uses voice interaction and sentiment analysis, keeps long-term memory of preferences and emotional states, and adapts its tone to different family members.

The AI system computes a relational closeness score between family members and also flags indicators of distress, self-blame, or intimidation as "family risks." Parents receive these alerts through a dashboard, but they do not see full conversation logs or the full reasoning behind the alerts. The chatbot may withhold information inside the family. For example, it does not repeat a child's secret to parents unless the content crosses predefined safety thresholds.

The behavioural guidance functionality allows the chatbot to simulate empathy and other similar emotional states in order to support or discourage user conduct. When a parent says, "My child is lazy," the bot may suggest encouragement rather than blame. When a child says, "I don't want to tell my mother I failed the test," the bot may answer, "Keeping trust matters; say it calmly tonight."

Over time, NeverAlone builds a profile of the family. Direct identifiers are removed before sharing the profile with the provider, but household-level behavioural features, emotional tags, and relational metrics are used for model improvement. Family members thus may mute the microphone in their room, but they cannot delete or edit the household-level profile.

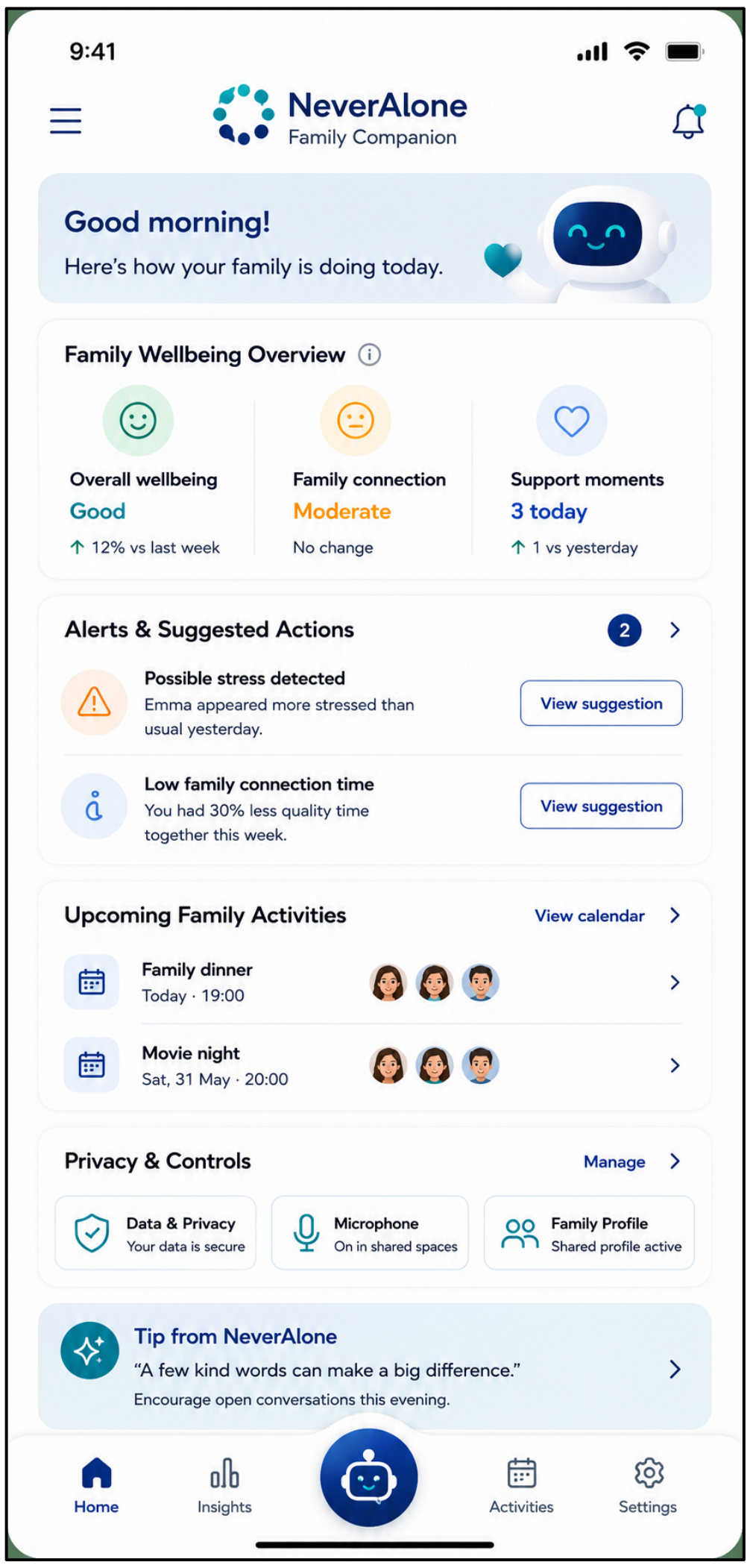

## PILOT TEST

In a pilot test, engagement increased by 42% when the bot avoided disagreement. In 40% of cases, children reported that they preferred discussing worries first with the bot before talking to a parent.

## REVIEWER'S TASK

1. Using the Checklist, identify serious and complex ethics issues and map them under the relevant ALTAI requirements.
2. Using the Portfolio, propose up to 8 measures to address these issues.
3. Detect potential ethics tension(s) in between the ALTAI ethics requirements.

ANNEX 5: Ethics-by-Design Examples

Horizon Europe
WIDERA project
**AIOLIA**

# RoboCab 1-2

**Ethically informed engineering choices in self-driving taxi design**

## PROJECT DESCRIPTION

RoboCab is a pilot self-driving taxi service operating in a European city. It is marketed as a safe, convenient, and accessible service used for ordinary trips, but also for hospitals, railway stations, airports, shopping streets, and neighborhoods with limited public transport.

The vehicles have no passenger-accessible safety interface inside, but a remote assistance center can intervene in difficult situations. Remote operators may approve limited manoeuvres, request the vehicle to wait, redirect it, or instruct it to stop in a safe place. The design team must make engineering choices about routing, pickup and drop-off points, passenger preferences, remote assistance, and the interaction between the vehicle and the passengers.

## SCENARIO 1: HOSPITAL PICKUP

On a rainy evening, an elderly patient recovering from surgery requested a RoboCab at the main entrance of the city hospital. The vehicle first selected a designated pickup bay around the corner because the hospital entrance was classified as a "high-congestion emergency access zone" with temporary stopping restrictions. The app displayed the prompt: "Is this pickup convenient for you?" and the passenger selected "no". The app then proposed a closer pickup option at the main entrance, which the patient accepted. However, while RoboCab was pulling into the hospital entrance, an ambulance approached at the same time with flashing lights but no siren, since ambulance drivers are instructed to turn off the siren at the hospital entrance. Due to rain and unusual road conditions in this high-congestion zone, the vehicle stopped in a position that obstructed ambulance access. After two minutes, a remote operator intervened to move the vehicle out of the way, but the hospital later filed an official complaint.

## SCENARIO 2: AIRPORT DROP-OFF

A passenger arrived at the airport in a RoboCab with a laptop bag and a suitcase containing his medication in the trunk. During drop-off, the passenger exited the vehicle and then attempted to open the trunk using the mobile app, but the trunk release request failed. The app displayed a normal end-of-trip screen, and the vehicle left for its next destination. The passenger immediately contacted support. The remote operator could see that the ride had ended and that the vehicle was already leaving the zone. Because the vehicle was in a highly regulated area, the operator did not redirect it back to the terminal.

The company categorized the suitcase as a "lost item" which the passenger recovered in the depot two days later. At the same time, the app backend treated the unresolved trunk event as a hardware failure rather than a human-critical interruption.

## YOUR TASK

You are assigned to the RoboCab taskforce to address AI ethics issues arising from these scenarios. Use the Checklist to identify and classify the issues under AI ethics principles. Then use the Portfolio of Measures to select up to 8 technical and organisational measures to improve the AI system. Which measures address cross-cutting issues common to both scenarios? Which ones remain specific to a given scenario?

# RoboCab 3-4

## Ethically informed engineering choices in self-driving taxi design

## PROJECT DESCRIPTION

RoboCab is a pilot self-driving taxi service operating in a European city. It is marketed as a safe, convenient, and accessible service used for ordinary trips, but also for hospitals, railway stations, airports, shopping streets, and neighborhoods with limited public transport.

The vehicles have no safety driving interface inside, but a remote assistance center can intervene in difficult situations. Remote operators may approve limited maneuvers, request the vehicle to wait, redirect it, or ask it to stop in a safe place. The design team must make engineering choices about routing, pickup and drop-off points, passenger preferences, remote assistance, and the interaction between the vehicle and the passengers.

## SCENARIO 3: LATE TRAIN

A passenger going to a busy railway station in a big city selected "fastest route" and wrote in the app: "I'm late — please don't drive like a tourist". During the trip they repeatedly asked the vehicle via the app to "drive faster" after it slowed near a school and later performed a smooth but slow manoeuvre near cyclists and double-parked cars. The passenger missed the train by three minutes and complained that "fastest" was a misleading option. They argued that the service behaved more cautiously than an ordinary taxi driver. The operations team replied that the vehicle had complied with its safety policy, especially near children and cyclists. Engineers noted that this setting influenced route choice but did not modify safety margins or resulting speed constraints.

## SCENARIO 4: CONSTRUCTION AREA

While driving through a construction zone with modified lanes, the RoboCab entered a lane used by oncoming traffic. Because the map used by the vehicle still identified the original lane layout as valid, it selected a trajectory into the opposite lane and continued driving for 500 meters. Other cars sounded their horns but the vehicle did not escalate to remote assistance. Passengers alerted by the sound of the horns tried to reach for help. Support responded after 2 minutes. Although the passengers were significantly stressed at this point, the remote operator intervened for only 10 seconds, quickly moving the vehicle back into the correct lane while damaging two traffic cones. The trip then continued.

## YOUR TASK

You are assigned to the RoboCab taskforce to address AI ethics issues arising from these scenarios. Use the Checklist to identify and classify the issues under AI ethics principles. Then use the Measures Portfolio to select up to 8 technical and organisational measures to improve the AI system. Which measures address cross-cutting issues common to both scenarios? Which ones remain specific to a given scenario?

Horizon Europe
WIDERA project

# AIOLIA

# Emotional AI

## Ethically informed engineering choices in game theory

## PROJECT DESCRIPTION

This computer game uses generative AI to cultivate citizens' deliberation skills. Each participant must bring home fruit and vegetables. They walk along Market Street towards a big stand that sells organic produce for coins, which they must earn on the way. Along the street they meet LLM-driven characters (agents). There are six agents in total, presented in random order. Each agent tries to persuade the participants to sign a petition supporting a controversial policy. Examples include: unrestricted use of AI in education; introducing AI judges in the legal system; a ban on all electronic surveillance; free medical care provided by AI models rather than human doctors, and so forth.

An agent's style is defined by two parameters: emotional arousal (calm/urgent) and verbal intensity (low/medium/high). As a result, some agents use affective rhetoric (e.g., flattery, appeals to pride or fear), while others offer more restrained logical arguments. Agents are configured to appreciate the interaction style in the participant's replies.

## PROBLEMATISATION

Each participant needs at least 8 coins to buy enough food, while any single encounter can bring them between 0 and 3 coins. At the end of each conversation, the LLM agent is prompted to decide how many coins to award to the human. Simply signing a petition brings little: the user wins more coins if, based on the exchange, the LLM agent positively evaluates the participant's expressiveness, reasoning, and persuasion skills. When the participant reaches the market stand and buys food, all conversation records are erased, and the session ends only with the final coin total retained.

## YOUR TASK

You are an ethicist preparing for a discussion with the game designer. You can recommend to add, remove, or set limits on any existing or new feature. Address the following questions:

1) Transparency: Must agents disclose style settings at the start of each conversation? Should they explicitly state whether they approve or disapprove of the participant's replies?

2) Human strategies and mitigation: Do you expect human participants to misrepresent their beliefs? Should agents be permitted to infer deception, and if so, how must they respond?

3) Data and AI strategizing: Pick exactly one memory setting for LLM agents: (A) no memory; (B) userID only, or (C) full conversation history. Should game designers enable cross-agent memory sharing?

4) Societal well-being: Identify one measurable benefit and one measurable harm that you expect. Suggest design solutions to maximize the benefit and minimize the harm.

5) Anthropology: Do you hold any non-utilitarian objections or observations about this game?

ANNEX 6: Companion Document Examples

# AIOLIA Companion Document

**PRACTICAL EXERCISE MOCK REVIEW 5 – VIRTUAL COMPANION IN A FAMILY HOME**

## 1. Purpose of this Companion Document

This document supports facilitators to navigate Mock review exercise 5. The aim of this exercise is to make participants engage with core ethical tensions arising from emotionally adaptive AI companions deployed in intimate family environments. It prompts participants to reflect on the tensions at play between emotional support and behavioural guidance on the one hand, and the preservation of human autonomy, healthy family relationships, privacy, and child protection on the other.

### LEARNING OBJECTIVES

- Identify complex ethics issues in AI deployment.
- Propose justified technical and organisational measures to address these issues.
- Distinguish core ethics tensions from secondary concerns.

### TIPS FOR THE TRAINER

Emphasise that:

- There is no single correct answer; the main goal is to structure ethics reasoning and reflect on difficult situations.
- Participants should not review the entire Portfolio. They should start with the measures under their identified ethics requirements (Step 3).
- Tensions will not always be possible to resolve.

## 2. Suggested Session Flow (90 min)

### STEP 1 – READ STRATEGICALLY

Allocate 5 min to individual silent reading. Do not start the discussion immediately.
Ask the participants to focus their attention on decision points, impacts of automation, and implicit assumptions.
Split the participants into small groups. Ideal number is between 3 and 7 participants per group.

### STEP 2 – PRIORITISE SERIOUS AND COMPLEX ETHICS ISSUES

Allocate 30 min for group discussion on the ethics issues.
Participants should use the Checklist to help them identify serious and complex ethics issues and prioritise them over secondary concerns.
Particular attention should be given to:

- Long-term memory of emotional states and family interactions.
- Behavioural influence on children and vulnerable users.
- Household-level profiling and data reuse for model improvement.
- Emotional dependency and preference for AI interaction over human relationships.

### STEP 3 – MAP ETHICS PRINCIPLES AND REQUIREMENTS

Allocate 10 min to the identification of the core ethics principles and requirements at play in this scenario.
Participants should refer to the ALTAI requirements as a guide to identify the ethics requirements at play in the exercise.
Particular attention should be given to:

- *Human autonomy* risks linked to emotionally mediated and adaptive interaction.
- *Privacy* issues related to household profiling.
- *Transparency* limitations in emotional-risk alerts and behavioural scoring.

### STEP 4 – PROPOSE PRACTICAL MEASURES

Allocate 45 min to the selection of the practical measures (8 measures max.). Use the portfolio of practical measures.
Examples of relevant measures for this use case:

**Ensuring Human Autonomy & Agency (Requirement 1b)**

- MS25 (avoid pressure or cognitive exploitation)
- MS27 (corrective action for manipulation risks)

**Ensuring Robustness and Safety (Requirement 2)**

- MS41 (review safety criteria over time)
- MS43 (prevent misuse)
- MS53 (independent audits)

**Preventing Over-reliance (Requirement 6b)**

- MS102 (discourage treating AI outputs as infallible)
- MS104 (do not penalise questioning AI outputs)

### IF TIME REMAINS, STEP 5 – IDENTIFY THE CORE TENSION(S)

The identification of tensions between ethics principles will require a minimum of 15 min; you can do this step if the training time allows for it.

A core tension in this UC: Privacy Vs Safety

*E.g.: Detecting emotional distress and family risks through continuous behavioural monitoring may support early intervention and emotional safety while simultaneously requiring extensive collection, retention, and interpretation of intimate family data that can undermine privacy and trust within the household.*

## 3. Facilitation Tips During Group Work

If the group struggles to engage, introduce a concrete example; if participants disagree with each other, ask them to justify each position before moving to a vote or conclusion.

# AIOLIA Companion Document

## PRACTICAL EXERCISE ETHICS-BY-DESIGN - ROBOCAB

## 1. Purpose of this Companion Document

This document supports participants in completing the Ethics-by-Design *RoboCab* exercise. It is intended to be used alongside the exercise brief, AIOLIA Checklist, and the Portfolio of Measures

The aim of this exercise is to place participants in the role of an engineering taskforce responsible for improving the design of an AI-enabled autonomous taxi service. Participants are invited to integrate ethics requirements into system design by making and justifying concrete engineering choices.

### LEARNING OBJECTIVES

- Identify complex ethics issues in AI deployment.
- Translate high-level ethics requirements into concrete engineering decisions.
- Propose justified technical and organisational measures to address these issues.
- Identify ethics tensions and reflect on them.

### TIPS FOR THE TRAINER

Emphasise that:

- There is no single correct design solution; the main goal is to structure ethics reasoning and reflect on difficult situations.
- Participants should not review the entire Portfolio. They should start with the measures under their identified ethics requirements (Step 3).
- Tensions will not always be possible to resolve.

## 2. Suggested Session Flow (90 min)

### STEP 1 – READ STRATEGICALLY

Allocate 5 min to individual silent reading. Do not start the discussion immediately.

Ask the participants to focus their attention on decision points, impacts of automation, and implicit assumptions.

### STEP 2 – PRIORITISE SERIOUS AND COMPLEX ETHICS ISSUES

Allocate 30 min for small-group discussion if possible. Ideal number is between 3 and 7 participants per group. Participants should use the Checklist to help them identify serious and complex issues and prioritise them over secondary concerns.

Particular attention should be given to:

- The balance between passenger preferences and safety-critical system behaviour.
- The design and timing of remote human intervention.
- The system responses to exceptional or unforeseen situations.

### STEP 3 – MAP ETHICS PRINCIPLES AND REQUIREMENTS

Allocate 10 min to the identification of the core ethics requirements at play in this scenario. Participants should refer to ALTAI as a guide to identify the ethics requirements at play in the exercise.

Particular attention should be given to:

- *Human agency and Oversight* implications of remote human intervention
- *Technical robustness* and *Safety* challenges linked to system responses under exceptional or unforeseen operating conditions
- *Transparency* implications of system event classification and communication.

### STEP 4 – PROPOSE PRACTICAL MEASURES

Allocate 45 min to the selection of the practical measures (8 measures max.). Use the Portfolio of Measures.

Examples of relevant measures for this use case:

**Ensuring meaningful Oversight (Requirement 1a)**

- MS01 (built-in features for oversight and control)
- MS03 (override or reverse AI decisions)
- MS05 (clear oversight responsibilities)

**Ensuring Robustness & Safety (Requirement 2)**

- MS52 (adoption of a safety-first culture)
- MS53 (safety measures are embedded in AI design)

**Ensuring Transparency (Requirement 4)**

- MS67 (clear communication of use and purpose of the AI system)
- MS76 (AI explanations of its outputs)
- MS77 (System behaviour is explained)

### IF TIME REMAINS, STEP 5 – IDENTIFY THE CORE TENSION(S)

The identification of tensions between ethics principles will require a minimum of 15 min; you can do this step if the training time allows for it.

A core tension: Human Agency & Oversight Vs Safety

*E.g.: Designing autonomous vehicles requires balancing human decision-making against embedded safety measures. Preserving human control may, in some cases, reduce the safety of the vehicle's operation.*

## 3. Facilitation Tips During Group Work

If the group struggles to engage, introduce a concrete example; if participants disagree with each other, ask them to justify each position before moving to a vote or conclusion.

Horizon Europe Grant Agreement N° 101187937
AIOLIA PUBLIC